\documentclass[10pt]{iopart}
\expandafter\let\csname equation*\endcsname\relax

\expandafter\let\csname endequation*\endcsname\relax
\usepackage{amsmath, amsthm, amssymb, amsfonts}
\usepackage{graphicx}
\usepackage{caption}
\usepackage{subcaption}
\usepackage{tabularx}
\usepackage{xfrac}
\usepackage{iopams}  
\usepackage{hyperref}
\hypersetup{colorlinks,
citecolor=blue, urlcolor=blue, linkcolor=blue}
\usepackage{upgreek}
\usepackage[greek,english]{babel}
\usepackage[utf8]{inputenc}
\usepackage{orcidlink}

\begin{document}
\title{Structure of Kerr black hole spacetimes in Weyl conformal gravity}

\author{Miguel Yulo Asuncion$^1$\orcidlink{0009-0003-5075-3107}, 
Keith Horne$^1$ \orcidlink{0000-0003-1728-0304}, Reinosuke Kusano$^1$\orcidlink{0000-0001-7569-7197}, Martin Dominik$^1$\orcidlink{0000-0002-3202-0343}}
\address{$^1$SUPA Physics and Astronomy, North Haugh, University of St\,Andrews, KY16\,9SS, Scotland,
United Kingdom}
\ead{md35@st-andrews.ac.uk}
\vspace{10pt}
\begin{indented}
\item[]September 2025
\end{indented}

\begin{abstract}
We analyze the stationary, uncharged, rotating, vacuum solution to Weyl conformal gravity. We elucidate the causal and ergoregion structure of the spacetimes found in the parameter space of the metric for positive mass. These are then compared to the analogous Kerr, Kerr-de Sitter, and Kerr-Anti-de Sitter solutions to general relativity. Additionally, we investigate and derive the extremal limits for both the horizons and ergosurfaces of the spacetimes. The horizon surface gravities and Hawking temperatures at the extremal horizon limits are then calculated to show that they vanish.

\vspace{1pc}
\end{abstract}

\submitto{\CQG}
%
%
%

\section{Introduction}\label{section:intro}
Albert Einstein's formulation of general relativity (GR) in 1915~\cite{Einstein:1915ca} revolutionized our understanding not just of gravity, but of the very nature of space and time. Over the decades, GR has successfully accounted for and predicted a plethora of physical phenomena. These include, among others, the precession of the perihelion of Mercury~\cite{Mercury}, the gravitational lensing of light~\cite{eclipse},  gravitational redshifts~\cite{poundrebka}, gravitational time dilation~\cite{dilation}, gravitational waves~\cite{abbott, hulsetaylor}, and the existence of black holes~\cite{oppenheimer, PenroseSingularity}. 

General relativity is governed by the Einstein field equations~\cite{Einstein:1915ca}
\begin{equation}\label{eq:einsteinfield}
    G_{\mu \nu} = 8 \uppi \  T_{\mu \nu},
\end{equation}
where $G_{\mu \nu}$ is the Einstein curvature tensor, and $T_{\mu \nu}$ is the stress-energy tensor.  We use geometrized units, where $G = c = \hbar = 1$, and a metric $g_{\mu \nu}$ signature of $(-, +, +, +)$. 

The impressive extent of physical processes accounted for by GR may seem to place its validity on firm footing. Contrary to this expectation, however, phenomena on quantum, galactic, intergalactic, and cosmological scales challenge its primacy. 

Observations of flat galactic rotation curves~\cite{flat, flat2} and anomalies in the dynamics of galaxy clusters~\cite{zwicky, roodcluster} have required positing the existence of invisible dark matter. Meanwhile, the discovery of the accelerating expansion of the universe~\cite{riessexpansion} has pointed towards the need for a dark energy component described by a cosmological constant $\Lambda$. The dominant cosmological paradigm, $\Lambda$CDM, includes such additional energy contributions in the Einstein field equations~\eqref{eq:einsteinfield}.

Serious qualms against the $\Lambda$CDM paradigm have arisen from the seemingly \textit{ad hoc} nature of the necessitated invocation of dark matter, a lack of direct detections of it~\cite{makingthecase}, and the vacuum energy density attributed to dark energy being $\sim 120$ orders of magnitude lower than what is predicted by particle physics theory~\cite{weinberg}.

The desire to resolve these tensions at astrophysical and cosmological scales motivates work into modified gravity theories such as Modified Newtonian dynamics (MOND)~\cite{MOND}, massive gravity~\cite{massivegravity}, Horndeski gravity~\cite{Bernardo_2019, Horndeski}, and others. 

It is also understood that GR breaks down at quantum mechanical scales. Attempting to use GR as a quantum field theory leads to non-renormalizable divergences at high energies~\cite{Bajardi_2021}. The failure of GR in the miscroscopic regime is evident in the presence of spacetime singularities, with divergent curvature scalars, in black hole models and at the dawn of the universe~\cite{HawkingSingularity}. 

The fundamental failings of GR at these ultraviolet (UV) scales calls for the formulation of viable quantum gravity theories. Current lines of work in this field include string theory~\cite{string}, loop quantum gravity~\cite{loopquantum}, twistor theory~\cite{twistor}, and a plethora of others. 

Weyl conformal gravity (CG)~\cite{Weyl1918} is one such alternative to general relativity. While the Einstein field equations~\eqref{eq:einsteinfield} possess an invariance to both coordinate $g_{\mu \nu}(x) \rightarrow g'_{\mu \nu} (x')$ and Lorentz $x^\mu \rightarrow \Lambda^\mu_\nu \ x^\nu$ transformations, CG derives from an invariance to local conformal transformations $g_{\mu \nu}(x) \rightarrow \Tilde{g}_{\mu \nu}(x) = \Omega^2(x) g_{\mu 
\nu}(x)$ as well~\cite{Turner_2020}. Here, $\Omega(x)$ is a conformal factor determining the extent of such stretching~\cite{VarieschiParameters1}. 

Requiring conformal invariance in the field equations points to a gravitational action of the form
\begin{equation}\label{eq:Weyl_action}
    I_{\rm W} = -\alpha_g \int \text{d}^4 x \sqrt{-g} \ C_{\lambda \mu \nu \kappa}C^{\lambda \mu \nu \kappa},
\end{equation}
where $\alpha_\text{g}$ is a dimensionless gravitational coupling constant, $g$ is the determinant of the metric $g_{\mu \nu}$, and $C_{\lambda \mu \nu \kappa}$ is the Weyl tensor given by the traceless part of the Riemann tensor $R_{\lambda \mu \nu \kappa}$~\cite{MannheimKazanas1989}. This can be obtained by subtracting all traces from the Riemann tensor
\begin{align}\label{eq:Weyl_tensor}
C_{\lambda \mu \nu \kappa} = R_{\lambda \mu \nu \kappa} -& \dfrac{1}{2}(g_{\lambda\nu} R_{\mu\kappa} - g_{\lambda\kappa}R_{\mu\nu} \notag - g_{\mu\nu}R_{\lambda\kappa}+g_{\mu\kappa}R_{\lambda\nu}) \notag \\&+ \dfrac{1}{6}{R^\alpha}_\alpha(g_{\lambda\nu}g_{\mu\kappa} - g_{\lambda\kappa}g_{\mu\nu}),
\end{align}
where, of course, $R_{\mu\nu} = g^{\rho \sigma} R_{\sigma \mu \rho \nu}$ is the Ricci tensor.

Varying the gravitational action $I_{\rm W}$~\eqref{eq:Weyl_action} with respect to the metric $g_{\mu \nu}$ gives~\cite{mannheimgammavalue}
\begin{equation}\label{eq:variation}
\frac{1}{\sqrt{-g}} \frac{\delta I_{\rm W}}{\delta g_{\mu \nu}} = - 2 \alpha_g W^{\mu \nu},
\end{equation}
where $W^{\mu \nu}$ is the traceless Bach tensor~\cite{Bach1921}. This may be written as
\begin{equation}\label{eq:bach_tensor} W_{\mu\nu}=W^{(2)}_{\mu\nu}-\dfrac{1}{3}W^{(1)}_{\mu\nu},
\end{equation}
with 
\begin{equation}
    \begin{aligned}
        \begin{cases}
            W^{(1)}_{\mu\nu}&=2g_{\mu\nu}{R^{;\lambda}}_{;\lambda}-2R_{;\mu;\nu}-2RR_{\mu\nu}+\dfrac{1}{2}R^2,\\
            W^{(2)}_{\mu\nu}&=\dfrac{1}{2}g_{\mu\nu}{R^{;\lambda}}_{;\lambda}+{{R_{\mu\nu}}^{;\lambda}}_{;\lambda}-{{R_{\mu}}^\lambda}_{;\mu;\lambda}-2R_{\mu\lambda}{R_{\nu}}^{\lambda}+\dfrac{1}{2}g_{\mu\nu}R_{\rho\lambda}R^{\rho\lambda}
        \end{cases}
    \end{aligned}
\end{equation}
as given in~\cite{MannheimKazanas1989}.

This then allows us to write the CG field equations in  a similar form to those for GR in~\eqref{eq:einsteinfield} as
\begin{equation}\label{eq:cgfield}
W_{\mu \nu}=\frac{1}{4 \ \alpha_\text{g}} \  T_{\mu \nu}.
\end{equation}
While $G_{\mu \nu}$ contains up to second-order derivatives of the metric, $W_{\mu \nu}$ has up to fourth-order derivatives, making CG a fourth-order theory~\cite{VarieschiParameters1, Bach1921}.

Some of the main departures of CG from GR may be identified by considering the static, uncharged, spherically symmetric vacuum solution to~\eqref{eq:cgfield}, found by Mannheim and Kazanas~\cite{MannheimKazanas1989}. We write this metric, which we refer to as CG Schwarzschild, as
\begin{equation}\label{eq:CG Schwarzschild}
\rmd s^2=-B(r) \ \rmd t^2+\frac{\rmd r^2}{B(r)}+r^2\left(\mathrm{~d} \theta^2+\sin ^2 \theta \ \rmd \phi^2\right).
\end{equation}
Taking the dimensionless mass parameter as $\beta = GM/c^2$ and the associated mass parameter~\cite{VarieschiFlyby}
\begin{equation}\label{eq:Mtwiddle}
    \widetilde{M} \equiv \beta\left(1-\frac{3}{2} \beta \gamma\right),
\end{equation}
we write the lapse function as
\begin{equation}\label{eq:CG B}
B(r)=1- 3 \beta \gamma -\frac{2\widetilde{M}}{r} +\gamma r-\kappa r^2.
\end{equation}
The parameters $\gamma $ and $\kappa $ are the additional constants that arise from CG being a fourth-order theory. 

Due to the fourth-order nature of CG, it remains an open question as to what corresponds to our more familiar notions of mass from GR. The conventional choice has been to consider $\beta$ (or $M$ in dimensionful units) as the inertial mass~\cite{makingthecase, mannheimgammavalue}, while $\widetilde{M}$~\eqref{eq:Mtwiddle} is some associated mass parameter~\cite{Brihaye_2009}. When we take $\gamma \rightarrow 0$, these two terms coincide $(\widetilde{M} = \beta)$, as we expect from GR. Additional difficulty in precisely defining what corresponds to the mass arises from CG metrics not generally being asymptotically flat, so that notions of ADM or Komar mass are not well-defined. However, since these distinctions do not play much of a role in our work, we do not explore these questions further, and simply refer to $\beta$ as the mass.

From CG Schwarzschild~\eqref{eq:CG Schwarzschild}, we recover the GR Schwarzschild solution by taking $\gamma, \kappa \rightarrow 0$. When solely $\gamma \rightarrow 0$, identifying $\kappa = \Lambda/3$ gives GR Schwarzschild-(Anti)-de Sitter for $\kappa > 0$ $(\kappa < 0)$.

The presence of the linear $\gamma r$ and quadratic $-\kappa r^2$ terms in the lapse function allows CG to account for flat galactic rotation curves~\cite{Mannheim2012Rot, Mannheim2013Rot, KeithRotation} and cosmological expansion~\cite{VarieschiParameters1, mannheimgammavalue, VarieschiCosmo} without dark matter or dark energy. With the $\gamma$ parameter being the coefficient of the linear $r$ term, it becomes relevant at galactic scales, and is thus mainly pertinent in explaining flat rotation curves, where dark matter is usually needed in $\Lambda$CDM. Meanwhile, since $\kappa$ is found in the quadratic $r^2$ term, it dominates the behavior at large $r$ cosmological scales, where dark energy is usually invoked. Readers interested in the nuances about to the roles and relations of the $\gamma$ and $\kappa$ parameters to such phenomena may be found in \cite{makingthecase, mannheimgammavalue}.  Fitting $\gamma$ and $\kappa$ to relevant phenomena and mass distributions remains an active field of research~\cite{VarieschiParameters1,mannheimgammavalue, VarieschiFlyby, VarieschiParameters2,  Varieschishadow}. 

Conformal gravity successfully replicates results of GR in such classical tests as perihelion precession and time dilation~\cite{sophielensing, Edery_1998} for small values of $\gamma$ and $\kappa$, making it a viable alternative to GR. 

Turning from these astrophysical and cosmological tensions to the UV regime, CG has also been considered as a candidate quantum gravity theory. Within the theory, for instance, conformal transformations have been shown to remove the divergence of curvature scalars at black hole singularities~\cite{Turner_2020, Bambikerrsing}. Furthermore, the fact that the coupling constant $\alpha_g$ in the action $I_{\mathrm{W}}$~\eqref{eq:Weyl_action} is dimensionless allows CG to be power counting renormalizable~\cite{makingthecase, mannheimgammavalue}, without necessitating higher dimensions like string theory requires.

One of the main criticisms of CG, as with other fourth-order theories, is that it should be subject to ghosts. However, it has been shown that a proper formulation of CG as a quantum theory is, in fact, ghostless~\cite{mannheim2007_1, bender2008}.

Despite these successes, there remain open problems in CG. These range from the getting the correct proportion of deuterium in big bang nucleosynthesis~\cite{nesbet}, properly modelling galaxy cluster dynamics~\cite{keithclusters, CGclusterbad}, to computing the decay process of binary pulsar orbits~\cite{makingthecase}.  

As in GR, black holes also arise in solutions to the CG field equations~\eqref{eq:cgfield}. Recent years have provided novel avenues for studying black holes in nature. These include the birth of gravitational wave observatories such as LIGO and VIRGO~\cite{abbott, Abbott2}, and even imaging from the Event Horizon Telescope (EHT)~\cite{EHT, EHTSgrA}. Such experiments open up the possibility of testing modified gravity theories with astrophysical black holes, making it a prime time to study black hole solutions in CG.

As astrophysical black holes are considered to be charge-free and rotating, the CG Kerr solution~\cite{MKmetric} is of marked interest. In Boyer-Lindquist coordinates~\cite{boyerlindquist}, 
with the spin parameter $a$ representing specific angular momentum of the mass, we have
\begin{equation}\label{eq:CG Kerr}
\begin{aligned}
\rmd s^2 & =-\left(1-\frac{2 \widetilde{M} r}{\rho^2}-k\left(r^2-a^2 \cos ^2 \theta\right)\right) \ \rmd t^2 \\
& +2\left(\frac{- 2 \widetilde{M} r a \sin ^2 \theta+k a\left(a^2\left(r^2+a^2\right) \cos ^4 \theta-r^4 \sin ^2 \theta\right)}{\rho^2}\right) \ \rmd t \ \rmd \phi \\
& +\left(\frac{\rho^2}{\Delta^{\mathrm{H}}}\right) \rmd r^2
+\left(\frac{\rho^2}{\Delta^{\theta}}\right) \ \rmd \theta^2  +\left(\frac{\Sigma^2\sin^2 \theta}{\rho^2} \right)  \ \rmd \phi^2 \ ,
\end{aligned}
\end{equation}
where we define the auxiliary functions~\cite{VarieschiFlyby}
\begin{equation}\label{eq:auxiliary}
\eqalign{
k\equiv\kappa+\frac{\gamma^2(1-\beta \gamma)}{(2-3 \beta \gamma)^2}\ , \cr
\rho^2 \equiv r^2 + a^2 \cos^2 \theta \ , \cr
\Delta^{\mathrm{H}} \equiv -k r^4 + r^2-2 \widetilde{M} r+a^2 \ , \cr
\Delta^{\theta} \equiv 1-k a^2 \cos ^2 \theta \cot ^2 \theta \ , \cr
\Sigma^2 \equiv \Delta^{\theta}\left(r^2+a^2\right)^2-a^2 \Delta^{\mathrm{H}} \sin ^2 \theta \ .}
\end{equation}

From this, we recover the GR Kerr metric~\cite{Kerr1963} for $\gamma, \kappa \rightarrow 0$. An appropriate conformal transformation and taking $\gamma \rightarrow 0$ 
reduces CG Kerr to GR Kerr-de Sitter (Kerr-dS) for $\kappa > 0$, and GR Kerr-Anti-de Sitter (Kerr-AdS) for $\kappa < 0$. Conformal transformations also show that CG Kerr~\eqref{eq:CG Kerr} is equivalent to CG Schwarzschild~\eqref{eq:CG Schwarzschild} when $a \rightarrow 0$~\cite{ VarieschiFlyby, MKmetric}. As such a transformation has not been explicitly shown in Boyer-Linquist coordinates in existing literature, we demonstrate this in~\ref{section: Appendix conformal}.

Thus far, the CG Kerr metric has been applied to studies of the flyby anomaly~\cite{VarieschiFlyby} and the shadow of Sagittarius A*~\cite{Varieschishadow}. However, the explicit structure of such CG Kerr spacetimes has yet to be studied. In this work, then, we explore the varying configurations of horizons and ergosurfaces in the parameter space of $\gamma$, $\kappa$, $\beta$, and $a$, as has been done for CG Schwarzschild~\cite{Turner_2020}. Furthermore, we elucidate the radial $(r)$ variation of causal and ergoregion structure. As the CG Kerr solution is an axisymmetric metric, we naturally evaluate these on the equatorial plane $\theta = \frac{\uppi}{2}$.

In performing such explorations, we also establish the means to find the extremal horizon limits of this metric, where distinct horizons coincide with one another. We also explore the extremal ergosurface limit, where the same occurs for ergosurfaces. Deviations from the GR Kerr extremal spin parameter of $a = \beta$ may allow for a test of CG from measurements of the spins of astrophysical black holes. Furthermore, such extremal limits are integral to some formulations of string theory and quantum gravity that depend on gauge-gravity dualities such as the Anti-de Sitter/Conformal Field Theory (AdS/CFT) correspondence~\cite{AdSCFT}.  These arise from the near-horizon geometries reducing to useful symmetries that permit Virasoro algebras, such as $\mathrm{AdS_2 \times S^2}$ for extremal GR Reissner-Nordstrom spacetimes~\cite{Kunduri_2013}.

We also calculate the surface gravities and the temperatures associated with Hawking radiation~\cite{Hawking1974, Hawking1975} for the horizons in these spacetimes. We check and confirm that these quantities indeed vanish at the extremal horizon limits, as one would expect from the GR counterparts to the CG Kerr solution. Explorations of the horizon thermodynamics of some other CG metrics have been done by~\cite{thermoliu, thermocorral}.

In this work, we explore the parameter space of the CG parameters $\gamma$ and $\kappa $, the mass $\beta $, and the spin parameter $a $. We consider both positive and negative values of $\gamma$ and $\kappa$. 

Values of $\gamma < 0$ had previously been said to be required to permit gravitational lensing~\cite{Edery_1998, Sultanagammavalue}, in contrast to $\gamma > 0$ values used to fit galaxy rotation curves and cosmological phenomena~\cite{ VarieschiParameters1, mannheimgammavalue, VarieschiParameters2}. However, more recent work~\cite{SultanaPositive, lensingpos}  has pointed out that erroneous approximations in calculating the bending angle in these previous studies of the lensing problem may have led to the conclusion that the wrong $(\gamma < 0)$ sign was needed. Thus, $\gamma > 0$ values have been found to align with both lensing and galactic rotation curve fits.

As we mentioned earlier, research to establish values for $\gamma$ and $\kappa$ by performing fits to rotation curves and cosmological expansion is still ongoing. Summarized in~\cite{VarieschiFlyby, Varieschishadow}, work by~\cite{mannheimgammavalue, VarieschiParameters1, VarieschiParameters2} give the orders of magnitude $\gamma \sim 10^{-30} - 10^{-28} \ \text{cm}^{-1}$ and $\kappa \sim 10^{-54} - 10^{-48} \ \text{cm}^{-2}$ when considering stars within galaxies as their sources. The solar system scale tests of perihelion precession~\cite{Sultanagammavalue}, time delay, and light deflection~\cite{sophielensing, Edery_1998} fall within the magnitudes of these constraints.  However, the specific values of such parameters may not necessarily be universal, and instead depend upon the details of the density distributions and structures of the sources being considered~\cite{Brihaye_2009}. Thus, as we are considering theoretical black holes in vacuum metrics in this study, we do not confine ourselves to any particular range of $\gamma$ and $\kappa$ values.

Since we are most interested in black hole solutions, we confine ourselves to positive mass $\beta > 0$. As a stationary metric, we have an invariance to a simultaneous transformation $\rmd t \rightarrow - \rmd t $ and $\rmd \phi \rightarrow - \rmd \phi $, so may then 
 simply consider spins $a > 0$. 

For simplicity, we shall recast all our coordinates and parameters in terms of $\beta$ to give the dimensionless forms $t/\beta \rightarrow t$, $r/\beta \rightarrow r$, $\beta \gamma \rightarrow \gamma$, $\beta^2 \kappa \rightarrow \kappa$, and $a/\beta \rightarrow a$. This is, of course, equivalent to taking $\beta = 1$. 

We begin, in section~\ref{section:geometry}, by introducing the equations governing the presence of ergosurfaces and horizons. Afterwards, in section~\ref{section:causal}, we establish the means to determine the causal structure and nature of horizons and ergosurfaces. The relevant equations for the extremal limits of CG Kerr spacetimes are derived in section~\ref{section:extremal}, and we also briefly explore the surface gravities and Hawking temperatures at these limits. We then move on, in section~\ref{section:parametric}, to establishing the general structure of CG Kerr spacetimes for $\kappa = 0$ along with elucidating the role of $\gamma$. The effects of varying  $\kappa$ and the spin $a$ are handled in sections~\ref{section:kappa} and~\ref{section:spin} respectively. We summarize our conclusions in section~\ref{section:conclusions}.

\section{CG Kerr geometry}\label{section:geometry}

\subsection{Ergoregions and ergosurfaces}\label{subsection:ergoregions}

While various authors define \textit{ergoregions} differently~\cite{kerrbrief}, we define them here as regions where $g_{tt} > 0$. Outside ergoregions, an observer may remain \textit{static} at fixed spatial coordinates $x^\mu = (t, r_0, \theta_0, \phi_0)$. Given that their four-velocity $u^\mu $ would only have a time component~\cite{Hobson_Efstathiou_Lasenby_2006}, and that for a massive particle $g_{\mu \nu}  u^\mu  u^\nu = -1$, such static observers may only exist where $g_{tt} < 0$. Within the ergoregions of rotating spacetimes, the frame-dragging effect~\cite{LenseThirring} forces trajectories to exhibit co-rotation with the mass~\cite{KerrdSPenrose}. In our classification, we denote ergoregions as $\mathrm{E}$ and non-ergoregions as $\mathrm{N}$. 

The ergosurfaces, where $g_{tt} = 0$, that bound ergoregions are then known as \textit{static limits}. On the equatorial plane, where $\theta = \frac{\uppi}{2}$, this condition is given by
\begin{equation}\label{eq:rescaled ergo}
\Delta^{\mathcal{E}} \equiv \left(-\rho^2 g_{tt}\right)\big|_{\theta = \frac{\uppi}{2}} =-k r^4 + r^2  - 2\widetilde{M}r = 0.
\end{equation}
Since $\cos \frac{\uppi}{2} = 0$, the spin $a$ dependent term in $g_{tt}$~\eqref{eq:CG Kerr} vanishes leaving $\Delta^{\mathcal{E}}$ with no explicit dependence on the spin $a$.

From this, we find that ergoregions $(\mathrm{E})$, where $g_{tt} > 0$, would have $\Delta^{\mathcal{E}} < 0$. Meanwhile, non-ergoregions $(\mathrm{N})$, where $g_{tt} < 0$, would have $\Delta^{\mathcal{E}} > 0$.

As a quartic with real coefficients, \eqref{eq:rescaled ergo} has four, two, or no real roots. Similar to GR, the CG Kerr solution possesses a ring singularity at $(r, \theta) = \left(0, \frac{\uppi}{2}\right)$. While real roots of  $\Delta^{\mathcal{E}}$ may be found for $r < 0$, we confine our work to discussions of features found for $r \geq 0$. A maximum of three ergosurfaces are found for $r \geq 0$.

Clearly, $\Delta^{\mathcal{E}}$ does not depend on spin $a$. There is also always a root and, therefore, an ergosurface $\mathcal{E}_0$, coincident with the ring singularity at $r = r^{\mathcal{E}}_0 = 0$. However, the interpretation of this ring singularity may deviate from that in GR, as it is argued that physical singularities may not exist in CG, due to conformal invariance. We refer the reader to the discussion in~\cite{Bambikerrsing}.

\subsection{Horizons}
In the Boyer-Lindquist coordinates that we use in~\eqref{eq:CG Kerr}, horizons or \textit{static limits}, meanwhile, are defined by where normal vectors $n_\mu$ to $r = \text{constant}$ surfaces become null. These occur when~\cite{Teukolsky_2015}
\begin{equation}
    n_\mu n_\nu \ g^{\mu \nu} = g^{rr} = \frac{\Delta^\mathrm{H}}{\rho^2} = 0,
\end{equation}
where  $\Delta^\mathrm{H}$ is as defined in~\eqref{eq:auxiliary}. As we can see, this is equivalent to finding where $g_{rr}$ switches sign as $1/g_{rr} = \Delta^\mathrm{H}/\rho^2= 0$. 

This switching of sign of $g_{rr}$ means that $r$ transitions from being a spacelike coordinate $(g_{rr} > 0)$ to a timelike one $(g_{rr} < 0)$. In finding such horizons then, we seek to solve 
\begin{equation}\label{eq:rescaled}
\Delta^{\mathrm{H}}= \Delta^{\mathcal{E}} + a^2 =- kr^4 + r^2  - 2\widetilde{M}r + a^2 = 0.
\end{equation}
We see that this quartic is nearly identical to the equation for the ergosurfaces~\eqref{eq:rescaled ergo}, but with an additional explicit dependence on the spin $a$.

To better orient ourselves, we review the ordering of ergosurfaces $\mathcal{E}$ and horizons $\mathrm{H}$ in some black hole spacetimes of GR. For GR Kerr and GR Kerr-AdS (with its attractive cosmological constant), we have the progression $\mathcal{E}_0 \rightarrow \mathrm{H}_{\mathrm{I}} \rightarrow \mathrm{H}_{\mathrm{O}}  \rightarrow \mathcal{E}_{\mathrm{O}}$ at radii $r^{\mathcal{E}}_0 < r^{\mathrm{H}}_{\mathrm{I}} < r^{\mathrm{H}}_{\mathrm{O}}< r^{\mathcal{E}}_{\mathrm{O}}$. Here, $\mathrm{H}_{\mathrm{I}}$ is the black hole's inner horizon, while $\mathrm{H}_{\mathrm{O}}$ is its outer horizon. $\mathcal{E}_{\mathrm{O}}$ is thus the outer ergosurface surrounding the outer horizon.

Black hole spacetimes of GR Kerr-dS, with a repulsive cosmological constant, generally have $\mathcal{E}_0 \rightarrow \mathrm{H}_{\mathrm{I}} \rightarrow \mathrm{H}_{\mathrm{O}}  \rightarrow \mathcal{E}_{\mathrm{O}} \rightarrow \mathcal{E}_{\mathrm{C}} \rightarrow \mathrm{H}_{\mathrm{C}}$ at radii $r^{\mathcal{E}}_0 < r^{\mathrm{H}}_{\mathrm{I}} < r^{\mathrm{H}}_{\mathrm{O}}< r^{\mathcal{E}}_{\mathrm{O}} < r^{\mathcal{E}}_{\mathrm{C}} < r^{\mathrm{H}}_{\mathrm{C}}$. Here, $\mathcal{E}_{\mathrm{C}}$ and $\mathrm{H}_{\mathrm{C}}$ are the cosmological ergosurface and horizon respectively.

It is worthwhile to here discuss the transitions between dS and AdS spacetimes. We may think of GR Kerr black hole spacetimes as possessing a cosmological horizon $\mathrm{H}_\mathrm{C}$ and ergosurface $\mathcal{E}_\mathrm{C}$ at $r = + \infty$, making them asymptotically flat. When we have the repulsive cosmological constant $\Lambda > 0$ of GR Kerr-dS spacetimes, these cosmological features ($\mathrm{H}_\mathrm{C}$ and $\mathcal{E}_\mathrm{C}$) are brought to finite $r$. For GR Kerr-AdS spacetimes, with $\Lambda < 0$, these cosmological features are pushed out \textit{beyond} $r = + \infty$.

Now, for CG Kerr, as with its ergosurface equation~\eqref{eq:rescaled ergo}, its quartic horizon equation~\eqref{eq:rescaled}  may have four, two, or no real roots. We shall see that a maximum of three of these roots will be found for $r > 0$, corresponding to $r^{\mathrm{H}}_{\mathrm{I}} < r^{\mathrm{H}}_{\mathrm{O}}< r^{\mathrm{H}}_{\mathrm{C}}$. 

\section{Establishing causal structure} \label{section:causal}
We consider the radial $(r)$ variation of causal structure in Boyer-Lindquist coordinates. We define \textit{timelike} $\mathrm{T}$ regions as being where $g_{rr} > 0$ and thus $\Delta^\mathrm{H} > 0$. Within such regions, motion to both increasing and decreasing $r$ is possible, as $r$ is \textit{spacelike}.

Correspondingly, in \textit{spacelike} $\mathrm{S}$ regions $r$ is a \textit{timelike} coordinate, so $g_{rr} < 0$ and thus $\Delta^\mathrm{H} < 0$. Within spacelike regions, world lines progress in only one direction in $r$. These come in two varieties.  In regions of the first type $\mathrm{S}^{-}$, all world lines progress toward decreasing $r$, as in the interiors of GR Schwarzschild black holes. In the second type $\mathrm{S}^{+}$, motion is possible only toward increasing $r$. This is now the type of spacelike region found exterior (at greater $r$) to cosmological horizons.

To distinguish between an $\mathrm{S}^{-}$ and an $\mathrm{S}^{+}$ region, once we know that $\Delta^\mathrm{H}< 0$, we look at the shape of the effective radial potential, and see whether trajectories are \textit{attracted} to smaller $r$ $(\mathrm{S}^-)$ or \textit{repelled} to larger $r$ $(\mathrm{S}^+)$. 

\subsection{Effective radial potential}\label{subsection: effective potential}

To find a radial effective potential, we begin with the equations of motion for a test particle in CG Kerr spacetimes as derived by~\cite{VarieschiFlyby, Varieschishadow}. These are 
\begin{equation}\label{eq: EOM}
\eqalign{
 \dot{t} = \frac{1}{\rho^2}\left(\frac{\left(r^2+a^2\right)\left[\left(r^2+a^2\right) E-a L_z\right]}{\Delta^\mathrm{H}}+\frac{a \sin ^2 \theta\left(L_z \csc ^2 \theta-a E\right)}{\Delta^{\theta}}\right), \cr
    \dot{r}^2 = \frac{\left[\left(r^2+a^2\right) E-a L_z\right]^2-\Delta^\mathrm{H}\left(\mathcal{Q}+\left(L_z-a E\right)^2 + \delta_1 r^2\right)}{\rho^4}, \cr
    \dot{\phi} = \frac{1}{\rho^2}
\left(\frac{a\left[\left(r^2+a^2\right) E-a L_z\right]}{\Delta^\mathrm{H}}+\frac{\left(L_z \csc ^2 \theta-a E\right)}{\Delta^{\theta}}\right), \cr
\dot{\theta}^2 = \left(\frac{\Delta^{\theta}}{\rho^2} \right)^2 p^2_\theta,
}
\end{equation}
where the dot represents differentiation with respect to an affine parameter $\sigma$, and $p_\theta$ is the $\theta$-component of the test particle four-momentum $p_\mu$. The other new quantities $(\delta_1, E, L_z, \mathcal{Q})$ derive from four corresponding conservation relations. The constancy of the norm of a particle's four-velocity $u^\mu$ may be expressed as
\begin{equation}
    |g_{\mu \nu} u^\mu u^\nu|^2 = \delta_1,
\end{equation}
with $\delta_1 = 0$ for massless particles, and $\delta_1 = 1$ for massive particles.

The time-translation invariance and axisymmetry yield the conserved constants $E$ and $L_z$ respectively. As CG Kerr spacetimes are not generically asymptotically flat, $E$ and $L_z$ do not directly represent the energy and angular momentum themselves. For simplicity, though, we still refer to $E$ and $L_z$ as ``energy" and ``angular momentum."

The fourth conserved quantity
\begin{equation}\label{eq: carter}
\mathcal{Q}=\Delta^{\theta} p_\theta^2+\frac{\left(a E \sin \theta-L_z \csc \theta\right)^2}{\Delta^{\theta}}-\left(L_z-a E\right)^2,
\end{equation}
is the CG version of Carter's constant, deriving from a Killing tensor~\cite{VarieschiFlyby, Carter}. 

Exploiting axisymmetry, we confine our derivation of a radial effective potential to the equatorial plane $(\theta = \frac{\uppi}{2})$, such that $p_\theta = 0, \rho^2 = r^2, \Delta^{\theta} = 1,$ and $\mathcal{Q} =0$. The radial equation of motion may then be written out as 
\begin{equation}
   r^2 \frac{\rm d r}{\rm d \sigma} = \pm \sqrt{\widetilde{R}(r)} \equiv \pm \sqrt{\left(\left(r^2+a^2\right) E-a L_z\right)^2-\Delta^\mathrm{H}\left(\delta_1 r^2 +\left(L_z-a E\right)^2 \right) }.
\end{equation}

Following the discussion of~\cite{StuchlikEquatorialKNdS}, we see that this equation presents us with a reality condition encoding the fact that test particle motion is only possible where $\widetilde{R}(r) \geq 0$. From this condition, we may solve for the energy $E$ of a massive test particle, where $\delta_1 = 1$. We get two solutions $ E \geq V_{+}(r)$ and $ E \leq V_{-}(r)$, where
\begin{equation}\label{eq: effective potential}
   V_{\pm}(r) \equiv \frac{a L_z\left(r^2 + a^2-\Delta^\mathrm{H}\right) \pm \sqrt{\Upsilon(r;a, L_z) }}{\left(r^2+a^2\right)^2-a^2 \Delta^\mathrm{H}},
\end{equation}
and where we have defined
\begin{equation}
    \Upsilon(r; a, L_z) \equiv \Delta^\mathrm{H}r^2\left[L_z^2 r^2 + \left((r^2 + a^2)^2  -\Delta^\mathrm{H} a^2 \right) \right]. 
\end{equation}

\begin{figure}
    \centering
    \begin{subfigure}[b]{0.48\textwidth}
        \includegraphics[width=\textwidth]{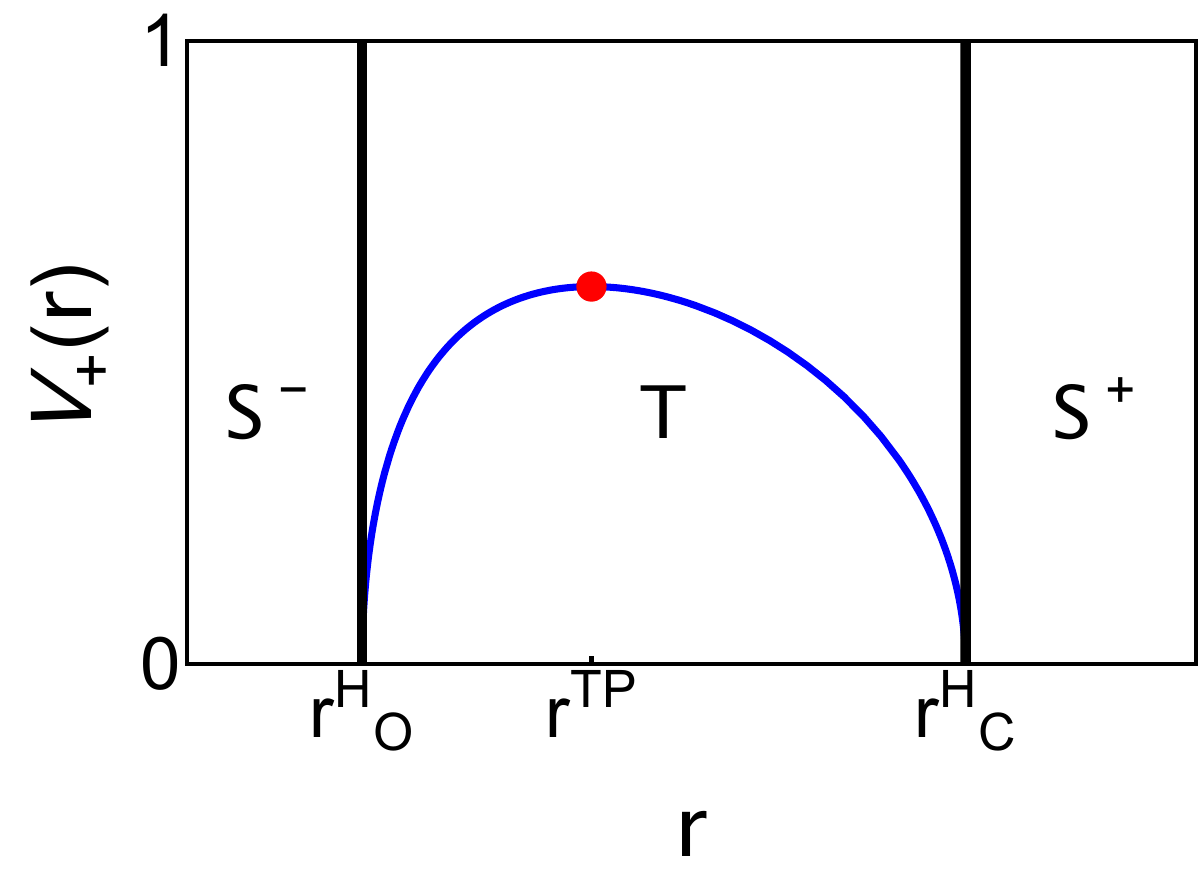}
        \caption[]%
        {$V_+(r)$}    
        \label{fig: V(r)}
    \end{subfigure}
    \hfill
    \begin{subfigure}[b]{0.49\textwidth}  
        \includegraphics[width=\textwidth]{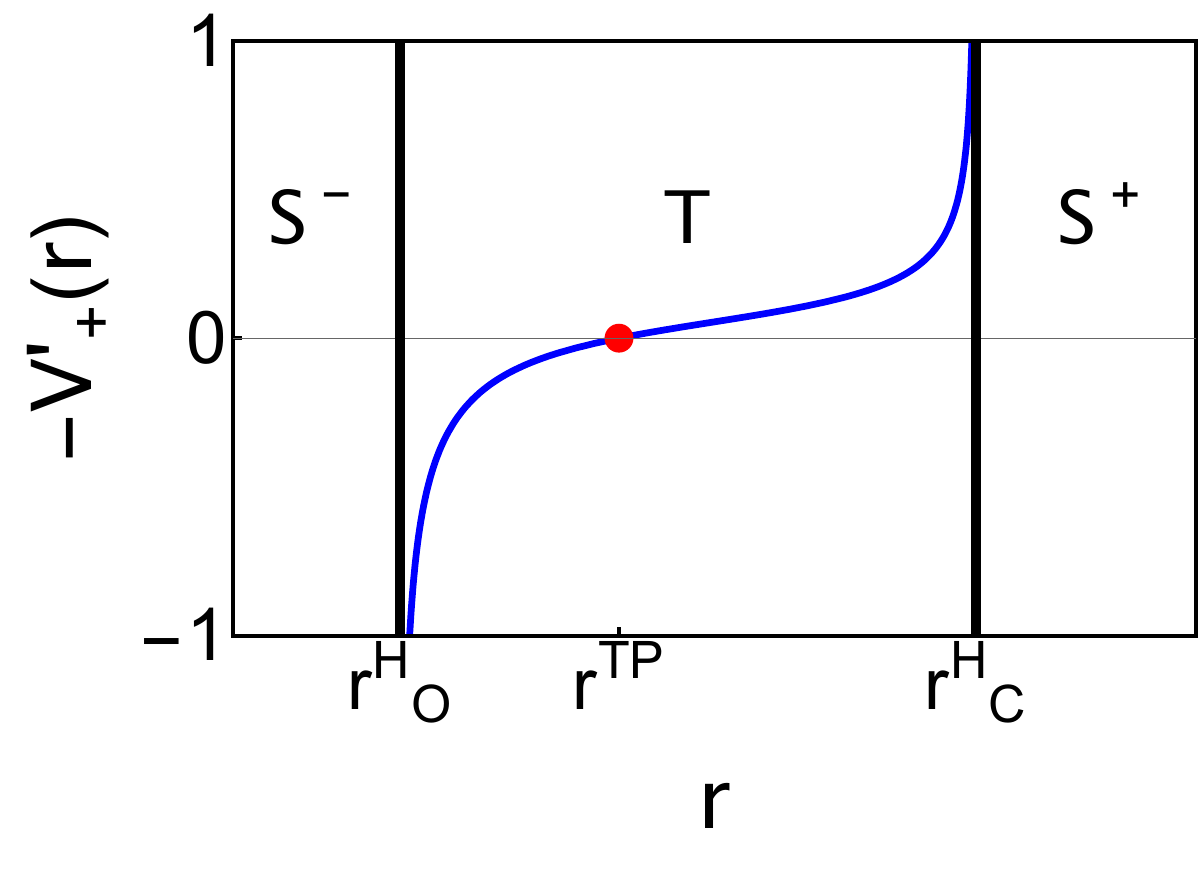}
        \caption[]%
        {$-V_+'(r)$}    
        \label{fig: V'(r)}
    \end{subfigure}

    \caption[]
    {Plots of $V_+(r)$ and $-V_+'(r)$, for $(\gamma, \kappa, a, L_z) = (0.1, 0.01, 0.25, 0)$, within a timelike $\mathrm{T}$ region. We have an $\mathrm{S}^-$ region to the left of the event horizon at $r^{\mathrm{H}}_\mathrm{O}$, and an  $\mathrm{S}^+$ region to the right of the cosmological horizon at $r^{\mathrm{H}}_\mathrm{C}$. The points (red) mark the values of these functions at the turning point of $V(r)$ at $r = r^{\mathrm{TP}}$.} 
    \label{fig:Effective Potential}
\end{figure}

As we are considering massive test particles, we may take the affine parameter to be the proper time, such that $\sigma \rightarrow \tau$. For future-directed trajectories $(\rmd t/\rmd \tau > 0)$, the positive root $ V_{+}(r)$ is the relevant one to be used~\cite{StuchlikEquatorialKNdS, MTW}.

To see whether a particle is attracted or repelled, we calculate the sign of $\Ddot{r} = -V_+'(r) \equiv - \rmd V_+(r)/\rmd r$. As our purpose here is merely to distinguish between $\mathrm{S}^-$ and $\mathrm{S}^+$ regions, for additional ease, we set a vanishing angular momentum $(L_z = 0)$. As it is rather unwieldy, we do not write out $-V'(r)$ in full here.

Due to the presence of the square root of $\Upsilon(r; a, L_z)$ in~\eqref{eq: effective potential}, $-V'_+(r)$ is not well-defined within $\mathrm{S}$ regions, where $\Delta^\mathrm{H} < 0$, and $-V'_+(r)$ becomes complex. This then ensures that any turning points of $V_+(r)$, where $V'_+ (r^{\mathrm{TP}}) = 0$, must occur within $\mathrm{T}$ regions. Figure~\ref{fig:Effective Potential} shows representative plots of $V_+(r)$ and $-V_+'(r)$ for parameter values $(\gamma, \kappa, a, L_z) = (0.1, 0.01, 0.25, 0)$.

If the horizon bounding an $\mathrm{S}$ region lies to the right of a turning point $(r^{\mathrm{TP}} < r^{\mathrm{H}})$, we calculate the sign of $-V_+'(r)$ at some $r$ where $r^{\mathrm{TP}} < r < r^{\mathrm{H}}$. Likewise, if the horizon lies to the left of a turning point $(r^{\mathrm{H}} < r^{\mathrm{TP}})$, then we calculate the sign of $-V_+'(r)$ at some $r$ where $r^{\mathrm{H}} < r < r^{\mathrm{TP}}$. If  $-V_+'(r) > 0$ ($-V_+'(r) < 0$), then a test particle is \textit{repelled} (\textit{attracted}), and we have an $\mathrm{S}^+$ ($\mathrm{S}^-$) region. 

This process of classifying $\mathrm{S}$ regions is clear from figure~\ref{fig: V'(r)}. Here, we see that $-V_+'(r) > 0$ to the right of the turning point $(r^{\mathrm{TP}} < r < r^{\mathrm{H}}_{\mathrm{C}})$, making the region immediately to the right of $r^{\mathrm{H}}_{\mathrm{C}}$ be $\mathrm{S}^+$. Correspondingly, $-V_+'(r) < 0$ to the left of the turning point $(r^{\mathrm{H}}_{\mathrm{O}} < r < r^{\mathrm{TP}})$, so the region immediately to the left of $r^{\mathrm{H}}_{\mathrm{O}}$ is $\mathrm{S}^-$.

\subsection{Classification of horizons and ergosurfaces}
We shall primarily encounter three types of horizons. Nomenclature varies within the field, so we outline our definitions here for clarity.

Firstly, \textit{Cauchy horizons} demarcate a transition from a $\mathrm{T}$ region to an $\mathrm{S}^-$ region as $r$ is increased $(\mathrm{T} \rightarrow \mathrm{S}^-)$. The inner horizons of GR Reissner-Nordstrom and GR Kerr black holes are of this type. We call these Cauchy horizons as they are found surrounding singularities in $\mathrm{T}$ regions. The causal past of an event within the $\mathrm{S}^-$ region outside a Cauchy horizon can be determined entirely by Cauchy information. Meanwhile, events within the $\mathrm{T}$ region inside the Cauchy horizon can be influenced by both Cauchy information and the singularity itself. Thus, such horizons separate regions from which Cauchy information is sufficient to causally determine a future event, while inside such a Cauchy horizon it may be insufficient~\cite{Podolsky}.

\textit{Event horizons} serve as the boundary of the transition $\mathrm{S}^-$ to $\mathrm{T}$ as $r$ is increased  $(\mathrm{S}^-\rightarrow \mathrm{T})$. Familiar examples are the horizons of GR Schwarzschild black holes and the outer horizons of GR Reissner-Nordstrom and GR Kerr black holes.

Lastly, \textit{cosmological horizons} separate $\mathrm{T}$ regions from $\mathrm{S}^+$ regions as $r$ is increased $(\mathrm{T} \rightarrow \mathrm{S}^+)$. Such cosmological horizons are found in spacetimes with de Sitter backgrounds. 

Both cosmological and Cauchy horizons can be thought of as being generated by an effective \textit{repulsive} effect. Cosmological horizons generate an $\mathrm{S}^+$ region exterior (at greater $r)$ to them. Meanwhile, Cauchy horizons can be thought to \textit{carve out} a $\mathrm{T}$ region within an otherwise fully $\mathrm{S}^-$ region. This \textit{repulsion} is how charge and spin in GR Reissner-Nordstrom and GR Kerr black holes generate an inner $\mathrm{T}$ region not present in GR Schwarzschild black holes.

The CG Kerr spacetimes we shall be exploring will mainly be characterized by the presence or absence of three horizons for $r \geq 0$. These are the inner horizon $\mathrm{H}_{\mathrm{I}}$, outer horizon $\mathrm{H}_{\mathrm{O}}$, and the outermost cosmological horizon $\mathrm{H}_{\mathrm{C}}$. 

We similarly label the three ergosurfaces by relation to these horizons. We have the ergosurface $\mathcal{E}_0$ at $r^\mathcal{E}_0 = 0$ mentioned earlier, the outer ergosurface $\mathcal{E}_{\mathrm{O}}$ found outside the outer horizon $\mathrm{H}_{\mathrm{O}}$, and the cosmological ergosurface $\mathcal{E}_{\mathrm{C}}$ located interior (at smaller $r$) to the cosmological horizon $\mathrm{H}_{\mathrm{C}}$. 

We then define \textit{Naked singularity} spacetimes as spacetimes with no event horizons $\mathrm{H}_\mathrm{E}$ protecting the singularity at the origin. In such naked singularity spacetimes, the only type of horizon that may be present is a cosmological one. This is similar to the naked singularities found in post-extremal GR spacetimes. These violate the Weak Cosmic Censorship Conjecture (WCCC). However, since the WCCC has not been proven in general even in GR~\cite{Wald1999}, we still explore such naked singularity spacetimes here. 

One may wonder where ergosurfaces are found in static spacetimes such as the Schwarzschild solutions in both GR and CG. In these, $g_{tt}$ and $g_{rr}$ switch signs at the same places, so horizons effectively serve as both the \textit{static} and \textit{stationary limit}. 

\section{Extremal limits and horizon temperatures}\label{section:extremal}

In this section, we develop the means to identify the locations and parameter values corresponding to extremal limits in the CG Kerr spacetimes. While extremality is most often looked at in the context of horizons, the fact that the horizons and ergosurfaces are distinct in Kerr spacetimes means that we shall also encounter extremal limits for ergosurfaces. More details for how expressions in this section are derived may be found in~\ref{section: Appendix quartics}.

Furthermore, we compute the surface gravities and Hawking temperatures of horizons in the CG Kerr spacetimes. We then confirm that these quantities vanish at the extremal horizon limits, as would be expected in GR.

\subsection{Extremal horizon limits}\label{subsection: extremal hor}
Our parameter space of $\gamma, \kappa,$ and $a$ permits three different extremal limits. The first of these is found when $\gamma = \frac{2}{3}$, for any values of $\kappa$ and $a$. This is due to the vanishing of the coefficient $(2- 3\gamma)$, giving rise to four coincident roots of the quartic $\Delta^\mathrm{H}$~\eqref{eq:rescaled}. 

This particular extremal limit then has four horizons at $r = 0$, the only time we have a fourth horizon for $r \geq 0$. We shall refer to this as the \textit{Empty} case, as we have no horizons for all $r > 0$. This is a rather peculiar spacetime, as $\Delta^\mathrm{H} = 0$ at $r = 0$, making the singularity null. We then have $\Delta^\mathrm{H} < 0$ for all $r > 0$, making all of it spacelike $(\mathrm{S})$. As this spacetime has no $\mathrm{T}$ region within $r \geq 0$, we cannot use the sign of $-V'_+(r)$~\eqref{eq: effective potential} to distinguish whether this is $\mathrm{S}^+$ or $\mathrm{S}^-$. However, as we shall see later on, the \textit{Empty} case is sandwiched between spacetimes in the parameter maps that only have cosmological horizons, so we can identify this as $\mathrm{S}^+$.

We are inclined to consider this particular spacetime to be ill-defined. It may be regarded as a degeneracy in the model, as $\gamma = \frac{2}{3}$ clearly leads to an infinity in $k$~\eqref{eq:auxiliary}. If some coordinate or conformal transformation can extricate this infinity, we have failed to discover it.

The second case is the \textit{extremal spin limit} $\mathrm{H_{\mathrm{ex(s)}}}$, where the inner Cauchy horizon $\mathrm{H}_{\mathrm{I}}$ and outer event horizon $\mathrm{H}_{\mathrm{O}}$ of the black hole merge $( r^\mathrm{H}_{\mathrm{I}} = r^\mathrm{H}_{\mathrm{O}} \equiv r^\mathrm{H}_{\mathrm{ex(s)}})$. For the GR Kerr spacetime, this limit occurs when the mass equals the spin $(a = 1)$. For GR Kerr-dS and GR Kerr-AdS spacetimes, the extremal spin value takes on other values $(a \neq 1)$ for different values of $\Lambda$~\cite{KerrdS}.

Thirdly, we have the \textit{extremal horizon cosmological limit} $\mathrm{H_{\mathrm{ex(c)}}}$. In this limit, the outer black hole event horizon $\mathrm{H}_{\mathrm{O}}$ and cosmological horizon $\mathrm{H}_{\mathrm{C}}$ coalesce  $( r^\mathrm{H}_{\mathrm{O}} = r^\mathrm{H}_{\mathrm{C}} \equiv r^\mathrm{H}_{\mathrm{ex(c)}})$. For GR Kerr-dS, this is known as the \textit{rotating Nariai limit}~\cite{dSHoedown}, as the near-horizon geometry can be shown to reduce to a rotating version of the cosmological Nariai metric~\cite{NariaiOriginal, NariaiHolo}. In this work, we do not show that the near-horizon geometry of this limit in CG Kerr spacetimes reduces to a CG corollary of the rotating Nariai spacetime. Therefore, we shall not use the term \textit{rotating Nariai} here.

We now derive the formulae to find these latter two limits. Going back to~\eqref{eq:rescaled}, we write the loci of horizons in terms of $\kappa$ as 
\begin{equation}\label{eq:kappa horizons}
    \kappa^{\mathrm{H}}(r ; \gamma, a)=\frac{\dfrac{\gamma^2\,(\gamma-1)}{(2-3 \,\gamma)^2} r^4+r^2-(2 -3\,\gamma) \,r+a^2}{r^4}.
\end{equation}
Extremal limits occur where horizons merge at extrema of $\kappa^{\mathrm{H}}$. 
We thus solve
\begin{equation}
0 = \frac{\partial \kappa^\mathrm{H}}{\partial r}
= \frac{ -2 \, r^2 
+ 3\,\left(2-3\,\gamma\right)\, r
- 4 \, a^2 } { r^5 }
\ ,
\end{equation}
to find
\begin{equation}
r_{\rm ex}^\mathrm{H}  
 = \frac{3\,\left(2-3\,\gamma\right) } { 4 }\,
 \left( 1 \pm 
 \sqrt{ 1 - 
 \frac{ 2\, a^2 }
 { 9 \, \left( 2 - 3 \, \gamma \right)^2 } } 
 \right)
 \ .
\end{equation}
 This gives the value of the spin $a$ at these extremal limits as
\begin{equation}\label{eq:a extremal}
a^{\mathrm{H}}_{\text{ex}}(r ; \gamma)= + \frac{1}{2} \sqrt{3(2- 3\gamma) r-2 r^2}.
\end{equation}
Since we have defined $a > 0$, we discard the negative root. Requiring that this be real, we get a restriction on $\gamma$ given by 
\begin{equation}\label{eq:gamma ex restric}
    \gamma^{\mathrm{H}}_{\text{ex}} \leq \frac{2}{3} - \frac{2}{9}r^{\mathrm{H}}_{\mathrm{ex}}.
\end{equation}
This tells us that the spin and horizon cosmological extremal limits occur at $\gamma$ values below the \textit{Empty} case at $\gamma = \frac{2}{3}$, since we have $r \geq 0$.

Applying conditions for repeated roots of quartic equations~\cite{quartic}, we require the discriminant of $\Delta^{\mathrm{H}}$~\eqref{eq:rescaled} to vanish $(D^{\mathrm{H}} = 0)$. Taking $k$ as defined in~\eqref{eq:auxiliary}, this reads
\begin{equation}\label{eq:hordisc}
D^{\mathrm{H}} \equiv -256a^6 k^3 - 128k^2a^4 - 16a^2k + (144a^2k + 4)(2-3\gamma)^2 k -27k^2(2-3\gamma)^4 .
\end{equation}
Now, when we set the values of two of the parameters from the set $(\gamma, \kappa, a)$,~\eqref{eq:hordisc} allows us to find the corresponding values of the third parameter that gives the extremal limits. For example, by setting specific values of $a$ and $\kappa$, we may solve for the extremal values $\gamma^{\mathrm{H}}_{\text{ex}}$. This may give us multiple values. However, when we apply the restriction in~\eqref{eq:gamma ex restric}, we shall get a maximum of two. The larger of these two will give the extremal spin case, and the smaller one the extremal horizon cosmological case $(\gamma^{\mathrm{H}}_{\text{ex(c)}} < \gamma^{\mathrm{H}}_{\text{ex(s)}})$. This reflects the fact that when we plot the locations of horizons in parameter maps of $\gamma$ for set values of $(\kappa, a)$, as we shall do in Section~\ref{section:parametric}, the extremal spin limit is found at a maximum, and the extremal horizon cosmological limit at a minimum.

After solving for $\gamma^{\mathrm{H}}_{\text{ex}}$, we can plug this and the value of $a$ that we have set into~\eqref{eq:a extremal} to find the location $r^{\mathrm{H}}_{\mathrm{ex}}$ where the corresponding horizons coincide. We may notice that this in fact gives two values of $r$. Only one of these is simultaneously a solution to both~\eqref{eq:kappa horizons} and~\eqref{eq:a extremal}, and thus the other value of $r$ is unphysical and may be discarded.

\subsection{Extremal ergosurface limits}\label{subsection:extremal ergo}
As the ergosurface quartic $\Delta^\mathcal{E}$~\eqref{eq:rescaled ergo} shows no spin $a$ dependence, we have no analogue of the extremal spin limit for ergosurfaces on the equatorial plane. This is also reflected in the fact that the innermost ergosurface $\mathcal{E}_0$ is always at $r^\mathcal{E}_0 = 0$.

Similar to the case of horizons, however, all four possible ergosurface roots coalesce at $r = 0$ for $\gamma = \frac{2}{3}$. Likewise, $\widetilde{\mathrm{\Delta}}^\mathcal{E} = 0$ at $r = 0$, and $\widetilde{\mathrm{\Delta}}^\mathcal{E} > 0$ for $r > 0$. 
Thus, in the \textit{Empty} spacetime all $r > 0$ is an ergoregion $(\mathrm{E})$, along with being $\mathrm{S}^+$.

The only other extremal case for ergosurfaces then is the \textit{extremal ergosurface cosmological limit} $\mathcal{E_{\mathrm{ex(c)}}}$  where the outer $\mathcal{E}_\mathrm{O}$ and cosmological $\mathcal{E}_\mathrm{C}$ ergosurfaces coincide $(r^\mathcal{E}_\mathrm{O} = r^\mathcal{E}_\mathrm{C} = r^\mathcal{E}_\mathrm{ex(c)})$.

As we did for the horizons, we may write the loci of ergosurfaces. After factoring out a power of $r$ giving the root $r^\mathcal{E}_0 = 0$,~\eqref{eq:rescaled ergo} reduces to a cubic, and we have 
\begin{equation}\label{eq: loci ergosurfaces}
    \kappa^\mathcal{E}(r ; \gamma)=\frac{\frac{\gamma^2(\gamma-1)}{(2-3 \gamma)^2} r^3+r-(2 -3\gamma) }{r^3}.
\end{equation}
Finding local extrema ($\partial \kappa^\mathcal{E}/\partial r = 0$) gives us the condition
\begin{equation}\label{eq: extremal ergosurface condition}
    \gamma^{\mathcal{E}}_{\mathrm{ex(c)}} = -\frac{2}{9}\left(r^{\mathcal{E}}_{\mathrm{ex(c)}}- 3\right).
\end{equation}

A cubic with real coefficients may have one or three real roots. The extremal ergosurface cosmological limit $\mathcal{E}_\mathrm{C}$ corresponds to where two of three real roots coincide. The condition for this~\cite{cubic} is that the discriminant of $\Delta^{\mathcal{E}}$~\eqref{eq:rescaled ergo}
\begin{equation}\label{eq:ergodisc}
    D^{\mathcal{E}} \equiv 4k - 27k^2 \left(2-3\gamma\right)^2
\end{equation}
vanishes $(D^{\mathcal{E}} = 0)$.
This can be manipulated to yield the condition
\begin{equation}\label{eq: UPS}
    \kappa = \frac{1 + 3\gamma}{27},
\end{equation}
which precisely corresponds to equation (12) in~\cite{Turner_2020}, giving the extremal horizon cosmological limit for the CG Schwarzschild case. This correspondence between the extremal horizon cosmological limit in the CG Schwarzschild case and the extremal ergosurface cosmological limit in the rotating CG Kerr case makes sense. Since, as discussed earlier in subsection~\ref{subsection:ergoregions}, we are operating on the equatorial plane $(\theta = \frac{\uppi}{2})$, $\Delta^{\mathcal{E}}$~\eqref{eq:rescaled ergo} has no dependence on $a$. Of course, in the non-rotating CG Schwarzschild spacetimes, the horizons and ergosurfaces are identical.

We then manipulate this to yield the extremal $\gamma$ value of
\begin{equation}\label{eq: ergo extremal gamma}
    \gamma^{\mathcal{E}}_{\mathrm{ex(c)}} =   9 \kappa - \frac{1}{3}.
\end{equation}
While we have not discovered a closed form expression for the $\gamma$ value at the extremal horizon cosmological limit $\mathrm{H}_{\mathrm{ex(c)}}$, the similarity in the forms of $\Delta^{\mathcal{E}}$~\eqref{eq:rescaled ergo} and $\Delta^{\mathrm{H}}$~\eqref{eq:rescaled} indicate that~\eqref{eq: ergo extremal gamma} may serve as a good approximation of it for small enough spin $a$.

\subsection{Horizon temperatures}\label{subsec:thermo}
We now compute the surface gravities and semiclassical Hawking temperatures~\cite{Hawking1974, Hawking1975} of the horizons found in the CG Kerr spacetimes.

As outlined in~\cite{HawkingradKNdS}, for a metric in Boyer-Lindquist coordinates, of the form~\eqref{eq:CG Kerr}, the surface gravity of the black hole event horizon $\mathrm{H}_{\mathrm{O}}$ is given by
\begin{equation}\label{eq: surface gravity BH}
    K_{\mathrm{O}} = \frac{1}{2} \frac{\partial_r \left[\Delta^\mathrm{H}\right]}{r^2 + a^2} \ \Bigg|_{r=r^\mathrm{H}_{\mathrm{O}}} = \frac{1}{2} \frac{-4k (r^{\mathrm{H}}_{\mathrm{O}})^3 + 2(r^{\mathrm{H}}_{\mathrm{O}})-2 \widetilde{M} }{(r^{\mathrm{H}}_{\mathrm{O}})^2 + a^2},
\end{equation}
where $k$ and $\widetilde{M}$ are as defined in~\eqref{eq:auxiliary}. From this, the Hawking temperature $T$ of this event horizon is then
\begin{equation}\label{eq: temp BH}
 T_{\mathrm{O}} = \frac{K_{\mathrm{O}}}{2 \uppi} = \frac{1}{4\uppi} \frac{-4k (r^{\mathrm{H}}_{\mathrm{O}})^3 + 2(r^{\mathrm{H}}_{\mathrm{O}})-2 \widetilde{M} }{(r^{\mathrm{H}}_{\mathrm{O}})^2 + a^2}.   
\end{equation}

Meanwhile, the respective formulas for the \textit{repulsive} Cauchy $\mathrm{H}_{\mathrm{I}}$ and cosmological $\mathrm{H}_{\mathrm{C}}$ horizons take on an overall negative sign to give
\begin{equation}\label{eq: surface gravity rep}
    K_{\mathrm{I/C}} = -\frac{1}{2} \frac{\partial_r \left[\Delta^\mathrm{H}\right]}{r^2 + a^2} \ \Bigg|_{r=r^\mathrm{H}_{\mathrm{{I/C}}}} = -\frac{1}{2} \frac{-4k (r^{\mathrm{H}}_{\mathrm{{I/C}}})^3 + 2(r^{\mathrm{H}}_{\mathrm{{I/C}}})-2 \widetilde{M} }{(r^{\mathrm{H}}_{\mathrm{{I/C}}})^2 + a^2},
\end{equation} 
and
\begin{equation}\label{eq: temp rep}
 T_{\mathrm{I/C}} = \frac{K_{\mathrm{I/C}}}{2 \uppi} = -\frac{1}{4\uppi} \frac{-4k (r^{\mathrm{H}}_{\mathrm{{I/C}}})^3 + 2(r^{\mathrm{H}}_{\mathrm{{I/C}}})-2 \widetilde{M} }{(r^{\mathrm{H}}_{\mathrm{{I/C}}})^2 + a^2}.   
\end{equation}

It is thus easy to see that at the extremal spin $\mathrm{H}_{\mathrm{ex(s)}}$ $(r^{\mathrm{H}}_{\mathrm{{I}}} = r^{\mathrm{H}}_{\mathrm{{O}}}= r^{\mathrm{H}}_{\mathrm{ex(s)}})$ and extremal horizon cosmological $\mathrm{H}_{\mathrm{ex(c)}}$ $(r^{\mathrm{H}}_{\mathrm{{O}}} = r^{\mathrm{H}}_{\mathrm{{C}}} = r^{\mathrm{H}}_{\mathrm{ex(c)}})$ limits the respective surface gravities and Hawking temperatures vanish as

\begin{eqnarray}\label{eq:temp extremal}
K_{\mathrm{ex(s)}} = K_{\mathrm{I}}(r^{\mathrm{H}}_{\mathrm{I}}=r^{\mathrm{H}}_{\mathrm{ex(s)}}) \ +  K_{\mathrm{O}}(r^{\mathrm{H}}_{\mathrm{O}}=r^{\mathrm{H}}_{\mathrm{ex(s)}})   = 0 \rightarrow T_{\mathrm{ex(s)}} = 0\\
K_{\mathrm{ex(c)}} =  K_{\mathrm{O}}(r^{\mathrm{H}}_{\mathrm{O}}=r^{\mathrm{H}}_{\mathrm{ex(c)}}) + K_{\mathrm{C}}(r^{\mathrm{H}}_{\mathrm{C}}=r^{\mathrm{H}}_{\mathrm{ex(c)}}) = 0 \rightarrow T_{\mathrm{ex(c)}} = 0.
\end{eqnarray}

We have thus shown that the horizon surface gravities $K$ and Hawking temperatures $T$ both vanish at the extremal horizon limits in CG Kerr spacetimes, just as we would expect for the GR Kerr and GR Kerr-(A)dS extremal horizon limits.

\section{General structure and role of $\gamma$}\label{section:parametric}

\begin{figure}[tb]
\centering
\includegraphics[width=0.7\linewidth]{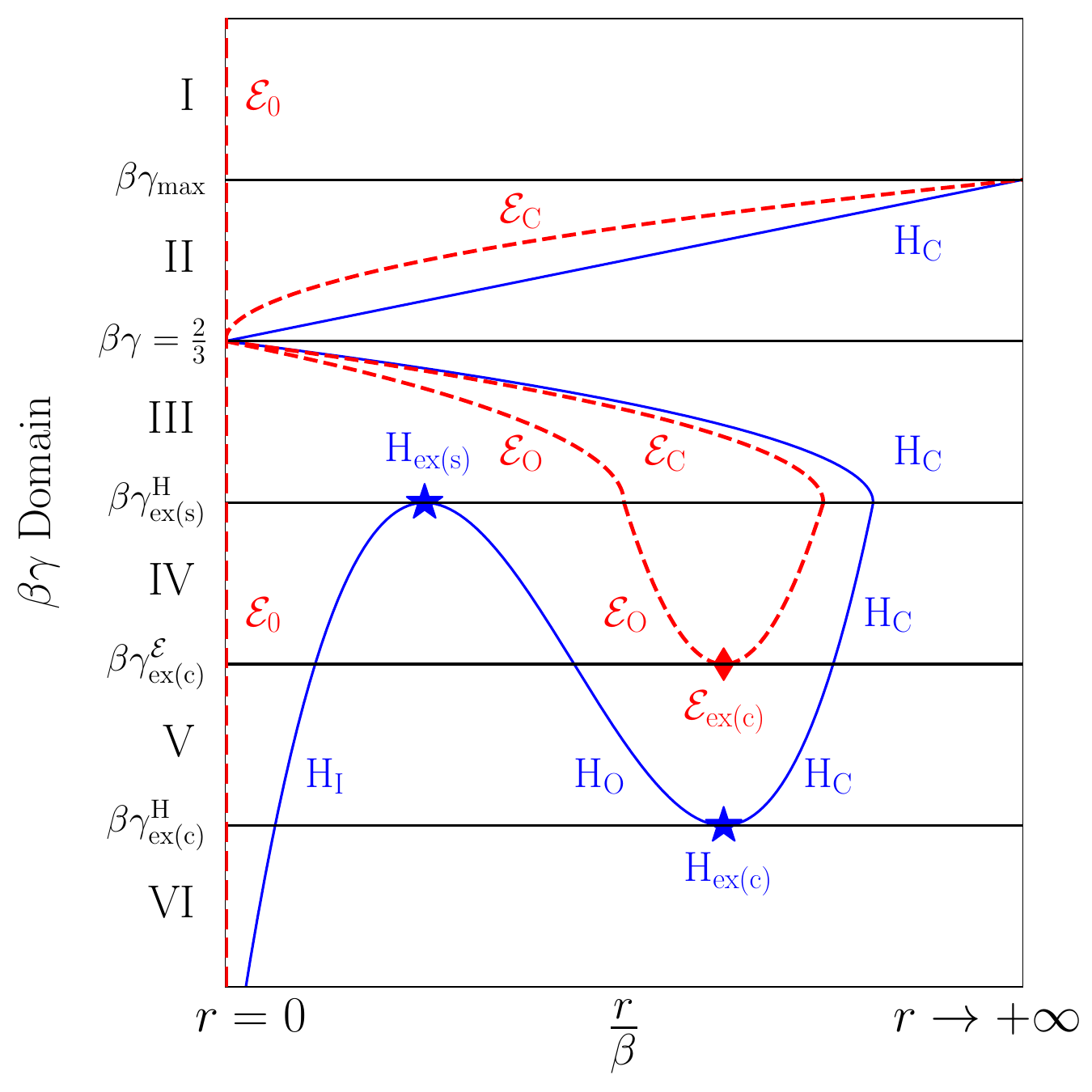}
\caption{Schematic diagram of a parametric plot $\beta \gamma$ vs $r/\beta$ showing the transitions between the domains in table~\ref{tab:domains}. Here, the horizons $\mathrm{H}$ are in blue and ergosurfaces $\mathcal{E}$ in dashed red. The extremal horizon limits are marked by stars, and the extremal ergosurface cosmological limit by a diamond. Note that distances between features have been exaggerated for clarity.}

\label{fig:CGK Domains}
\end{figure}

\begin{figure}
    \centering
    \begin{subfigure}[b]{0.445\textwidth}
        \includegraphics[width=\textwidth]{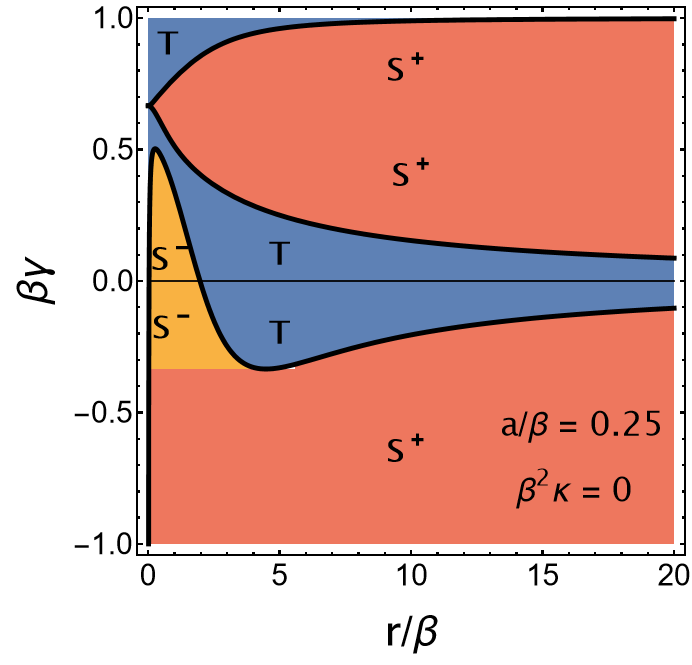}
        \caption[]%
        {Causal structure}    
        \label{fig: general causal1}
    \end{subfigure}
    \hfill
    \begin{subfigure}[b]{0.445\textwidth}  
        \includegraphics[width=\textwidth]{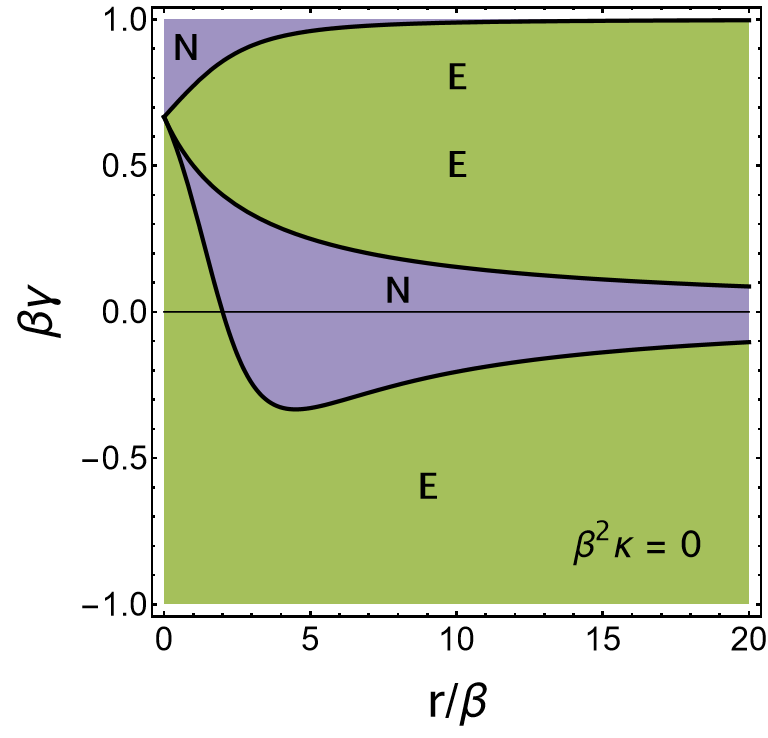}
        \caption[]%
        {Ergoregion structure}    
        \label{fig: general ergo1}
    \end{subfigure}
    \hfill
    \begin{subfigure}[b]{0.7\textwidth}  
        \includegraphics[width=\textwidth]{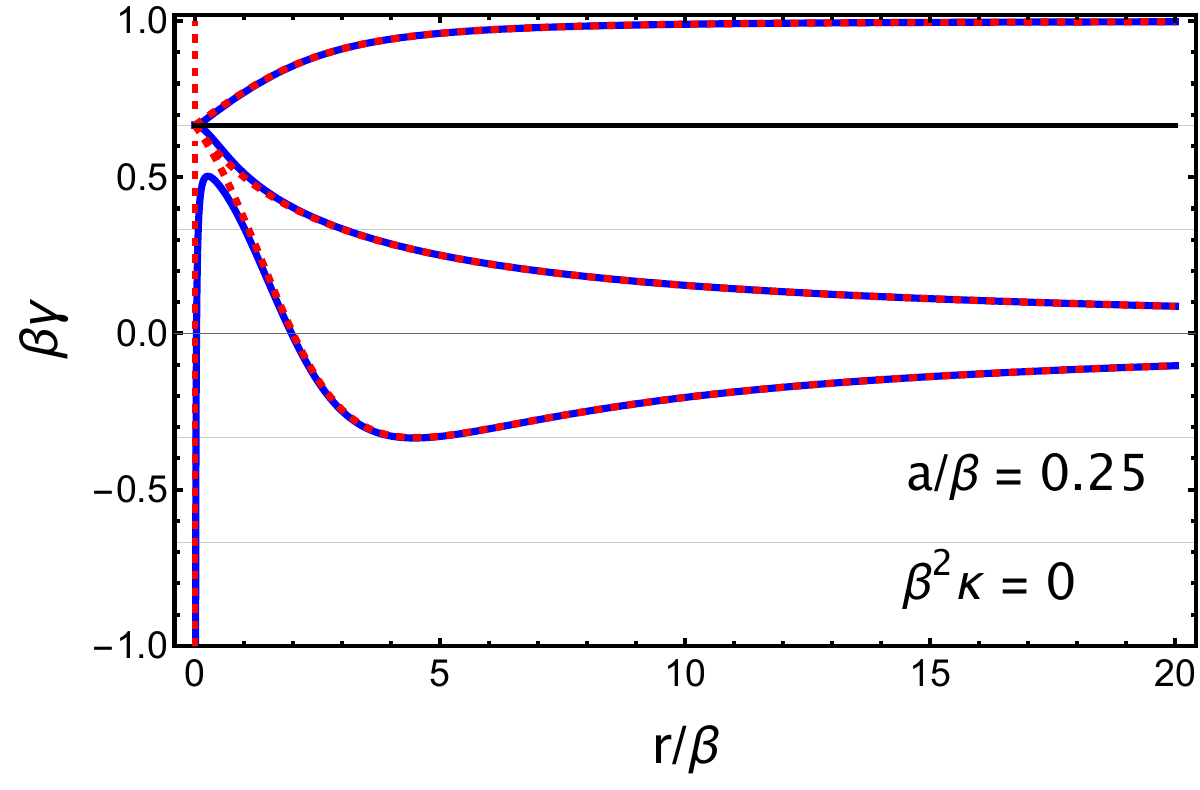}
        \caption[]%
        {Horizons (blue) and ergosurfaces (dashed red)}    
        \label{fig: general features}
    \end{subfigure}

    \caption[]
    {Causal structure, ergoregion structure, and plot of horizons (blue) and ergosurfaces (dashed red) of the CG Kerr spacetime for $(\beta^2\kappa, a/\beta) = (0, 0.25)$. The solid black line represents the \textit{Empty} case at $\beta \gamma = \frac{2}{3}$.} 
    \label{fig:General Structure}
\end{figure}

\begin{table}

\caption{\label{tab:domains} Domains of the parametric maps of spacetime features and regions that may be found for $\kappa \geq 0$. For $\kappa < 0$, we see these spacetimes supplemented by additional domains described in table~\ref{tab:domains neg}. Here, $\gamma^{\mathrm{H}}_{\mathrm{ex(s)}}$, $\gamma^{\mathrm{H}}_{\mathrm{ex(c)}}$, and $\gamma^{\mathcal{E}}_{\mathrm{ex(s)}}$ are the values of $\gamma$ at the extremal horizon spin limit, extremal horizon cosmological limit, and extremal ergosurface cosmological limit respectively. 
}

\begin{tabular}{@{}lllll}
\br
\rm Label &Domain & Features Present & Causal Structure & Ergoregions\\
\mr

I&$\gamma >\gamma_{\text{max}}$ &$\mathcal{E}_0$ & $\mathrm{T}$ & $\mathrm{N}$\\

II&$\frac{2}{3} <  \gamma \leq \gamma_{\text{max}}$ & $\mathcal{E}_0 , \mathcal{E}_\mathrm{C}, \mathrm{H}_\mathrm{C}$ & $\mathrm{T} \rightarrow \mathrm{S}^+$ & $\mathrm{N} \rightarrow \mathrm{E}$\\

\textit{Empty}&$\gamma = \frac{2}{3}$&All at $r = 0$ & $\mathrm{S}^+$ & $\mathrm{E}$\\

III& $\gamma^{\mathrm{H}}_{\mathrm{ex(s)}} <  \gamma < \frac{2}{3}$ &$\mathcal{E}_0 , \mathcal{E}_\mathrm{O} , \mathcal{E}_\mathrm{C} , \mathrm{H}_\mathrm{C}$ & $\mathrm{T} \rightarrow \mathrm{S}^{+}$ & $\mathrm{E} \rightarrow \mathrm{N} \rightarrow \mathrm{E}$\\

$\mathrm{H}_{\mathrm{ex(s)}}$& $\gamma = \gamma^{\mathrm{H}}_{\mathrm{ex(s)}}$ &$\mathcal{E}_0 , \mathrm{H}_\mathrm{I} = \mathrm{H}_\mathrm{O},\mathcal{E}_\mathrm{O} , \mathcal{E}_\mathrm{C} , \mathrm{H}_\mathrm{C}$ & $\mathrm{T} \rightarrow \mathrm{S}^- \rightarrow \mathrm{T} \rightarrow \mathrm{S}^+ $ & $\mathrm{E} \rightarrow \mathrm{N} \rightarrow \mathrm{E}$\\

IV & $\gamma^{\mathcal{E}}_{\mathrm{ex(c)}} <  \gamma < \gamma^{\mathrm{H}}_{\mathrm{ex(s)}}$ & $\mathcal{E}_0 , \mathrm{H}_{\mathrm{I}}, \mathrm{H}_{\mathrm{O}},  \mathcal{E}_{\mathrm{O}}, \mathcal{E}_\mathrm{C}, \mathrm{H}_\mathrm{C}$ & $\mathrm{T} \rightarrow \mathrm{S}^- \rightarrow \mathrm{T} \rightarrow \mathrm{S}^+$ & $\mathrm{E} \rightarrow \mathrm{N} \rightarrow \mathrm{E}$\\

$\mathcal{E}_{\mathrm{ex(c)}}$ & $\gamma =\gamma^{\mathcal{E}}_{\mathrm{ex(c)}}$ & $\mathcal{E}_0 , \mathrm{H}_{\mathrm{I}} , \mathrm{H}_{\mathrm{O}} ,  \mathcal{E}_{\mathrm{O}} = \mathcal{E}_\mathrm{C}, \mathrm{H}_\mathrm{C}$ & $\mathrm{T} \rightarrow \mathrm{S}^- \rightarrow \mathrm{T} \rightarrow \mathrm{S}^+$ & $\mathrm{E} \rightarrow \mathrm{N} \rightarrow \mathrm{E}$\\

V & $\gamma^{\mathrm{H}}_{\mathrm{ex(c)}}< \gamma < \gamma^{\mathcal{E}}_{\mathrm{ex(c)}}$ & $\mathcal{E}_0 , \mathrm{H}_{\mathrm{I}} , \mathrm{H}_{\mathrm{O}} ,   \mathrm{H}_\mathrm{C}$ & $\mathrm{T} \rightarrow \mathrm{S}^- \rightarrow \mathrm{T} \rightarrow \mathrm{S}^+$ & $\mathrm{E} $\\

$\mathrm{H}_{\mathrm{\mathrm{ex(c)}}}$ & $\gamma = \gamma^{\mathrm{H}}_{\mathrm{ex(c)}}$ & $\mathcal{E}_0 , \mathrm{H}_{\mathrm{I}} , \mathrm{H}_{\mathrm{O}} =   \mathrm{H}_\mathrm{C}$ & $\mathrm{T} \rightarrow \mathrm{S}^- \rightarrow \mathrm{T} \rightarrow \mathrm{S}^+$ & $\mathrm{E} $\\

VI & $\gamma < \gamma^{\mathrm{H}}_{\mathrm{ex(c)}}$ & $\mathcal{E}_0 , \mathrm{H}_{\mathrm{I}} $ & $\mathrm{T} \rightarrow \mathrm{S}^+$ & $\mathrm{E} $\\ 

\br
\end{tabular}

\end{table}

We first consider the case of $\kappa = 0$ to get a general picture of the structure of CG Kerr spacetimes on the equatorial plane $(\theta = \frac{\uppi}{2})$. This is instructive since a dependence on $\gamma$ is one of the most distinctive features of CG solutions, as the de Sitter term being linearly dependent on a parameter such as $\Lambda$ is of course present in GR (A)dS metrics. We shall see the effect of the variation of $\kappa$ and spin $a$ later.

To achieve this, we generate parametric plots with the parameter $\gamma$ on the vertical axis and the radial coordinate $r$ on the horizontal axis. We choose $\gamma$ to construct our plots with as $\Delta^\mathcal{E}$~\eqref{eq:rescaled ergo} and $\Delta^\mathrm{H}$~\eqref{eq:rescaled} have the most complicated dependence on $\gamma$. Note that we have restored factors of $\beta$ in the figures for clarity, but we shall maintain our use of the dimensionless quantities in the body of the text.

 We summarize features and regions present for $\kappa \geq 0$ in table~\ref{tab:domains}. A schematic of a parametric plot of $\gamma$ against $r$ clearly showing these domains is presented in figure~\ref{fig:CGK Domains}. As we shall see in subsequent sections, the effects of varying $\kappa$ and $a$ will be to shift or entirely remove some of these domains from our maps.

The plots of causal structure, ergoregion structure, and spacetime features (horizons and ergosurfaces) for the parameter values $( \kappa, a) = (0, 0.25)$ are shown in figure~\ref{fig:General Structure}.

When we have $\kappa = 0$, we recover the GR Kerr case when $\gamma = 0$ as well, and thus $k = 0$~\eqref{eq:auxiliary}. The cosmological features $\mathcal{E}_{\mathrm{C}}$ and $\mathrm{H}_{\mathrm{C}}$, and the corresponding $\mathrm{E}$ and $\mathrm{S}^+$ regions they bound would then be found at $r = + \infty$, as discussed earlier. This reflects the fact that it is $k$, from the quartic de Sitter term in $\Delta^\mathrm{H}$~\eqref{eq:rescaled} that governs the causal structure of the background. We expect de Sitter backgrounds for $k > 0$ and Anti-de Sitter backgrounds for $k < 0$. The cosmological ergosurfaces and ergoregions would of course follow suit from the similar role of $k$ in $\Delta^\mathcal{E}$~\eqref{eq:rescaled ergo}. 

Now, we briefly describe the key features of each domain. Black hole spacetimes, containing an event horizon $\mathrm{H}_\mathrm{C}$, are described by the domains $\mathrm{H}_{\mathrm{ex(s)}}$, IV, $\mathcal{E}_{\mathrm{ex(c)}}$, V, and $\mathrm{H}_{\mathrm{ex(c)}}$. All other domains are naked singularity spacetimes, or the degenerate \textit{Empty} case. We find that the singularity is timelike in all domains, except the \textit{Empty} case where it is null.

Starting with the horizonless Domain I, where $\gamma > \gamma_{\text{max}}$. We see that the only feature present is the ergosurface $\mathcal{E}_0$ at $r = 0$. The whole of $r > 0$ is both timelike $(\mathrm{T})$ and a non-ergoregion $(\mathrm{N})$. Domain I thus has a naked singularity at $r = 0$.

The value $\gamma_{\text{max}}$ is found by solving for where the respective horizon $D^{\mathrm{H}}$~\eqref{eq:hordisc} and ergosurface $ D^{\mathcal{E}}$~\eqref{eq:ergodisc} discriminants simultaneously vanish. This occurs when $k = 0$~\eqref{eq:auxiliary}.

For our case here, where $\kappa = 0$, we can define
\begin{equation}\label{eq: k0}
    k_0 \equiv k(\kappa = 0) = \frac{\gamma^2(1-\gamma)}{(2-3\gamma)^2}.
\end{equation}
We thus find $\gamma_{\text{max}} = 1$ from $k = k_0 = 0$. The value of $\gamma_{\text{max}}$ would of course vary when we include $\kappa$. Thus, as Domain I lies above $\gamma_{\text{max}}$ such that $k < 0$, we may consider Domain I as having an Anti-de Sitter background.

In Domain II then, we  see the cosmological horizon $\mathrm{H}_{\mathrm{C}}$ and ergosurface $\mathcal{E}_{\mathrm{C}}$ curves asymptotically approach $\gamma_{\text{max}} = 1$. Here, $\mathcal{E}_\mathrm{C}$ is the boundary for the transition in ergoregion structure $\mathrm{N} \rightarrow \mathrm{E}$ as $r$ is increased. Likewise, the cosmological horizon $\mathrm{H}_{\mathrm{C}}$ demarcates the transition from a timelike region to a cosmological spacelike region $(\mathrm{T} \rightarrow \mathrm{S}^+)$. At precisely $\gamma = \gamma_{\text{max}}$, these cosmological features ($\mathrm{H}_\mathrm{C}$ and $\mathcal{E}_\mathrm{C}$) are found at $r = + \infty$. As the black hole event horizon is absent, the singularity in these spacetimes is naked. 

As touched upon earlier, the \textit{Empty} case occurs at $\gamma = \frac{2}{3}$. Here, the singularity at $r = 0$ is null $(\Delta^\mathrm{H} = 0)$, and the four horizons coalesce here. Since $k = + \infty$, and thus $k > 0$, we may consider the \textit{Empty} case as having a de Sitter background, and thus being $\mathrm{S}^+$, along with being $\mathrm{E}$ for all $r > 0$. Since all four possible horizons are coincident with the singularity, we refrain from considering it as naked.

Domain III shares an identical causal structure $(\mathrm{T} \rightarrow \mathrm{S}^+)$ to Domain II, and thus the singularity is naked. However, the appearance of the outer ergosurface $\mathcal{E}_\mathrm{O}$ means that the ergoregion structure is now $\mathrm{E} \rightarrow \mathrm{N} \rightarrow \mathrm{E} $.  

The extremal spin case $\mathrm{H}_{\mathrm{ex(s)}}$
marks the upper limit of black hole spacetimes, and thus lies on a local maximum of the blue horizon curve. It has the same ergoregion structure as Domain III. At this limit, the inner Cauchy $\mathrm{H}_{\mathrm{I}}$ and outer event $\mathrm{H}_{\mathrm{O}}$ horizons of the black hole appear as a double root at $r = r^\mathrm{H}_{\mathrm{ex(s)}}$. Its causal structure is now $\mathrm{T} \rightarrow \mathrm{S}^- \rightarrow \mathrm{T} \rightarrow \mathrm{S}^+ $. The $\mathrm{S}^-$ region lies between the inner Cauchy $\mathrm{H}_{\mathrm{I}}$ and outer event $\mathrm{H}_{\mathrm{O}}$ horizons. The existence of this region may not make immediate sense, given that the extremal case has these two horizons at the same radial coordinate $r = r^{\mathrm{H}}_{\text{ex(s)}}$. However, it has been found that extremal black holes still possess a non-zero volume between the horizons~\cite{Carrollextremal}.

We find the bulk of the black hole spacetimes within Domain IV, for $\gamma$ values between the extremal spin $\gamma^{\mathrm{H}}_{ex(s)}$ and extremal ergosurface cosmological $\gamma^{\mathcal{E}}_{ex(c)}$ limits. Domain IV shares the causal and ergoregion structure of the extremal spin case. The Domain IV spacetimes all have a Kerr black hole with a $\mathrm{T}$ region encased by the inner Cauchy horizon $\mathrm{H}_\mathrm{I}$, followed by an $\mathrm{S}^-$ region between $\mathrm{H}_\mathrm{I}$ and the outer event horizon $\mathrm{H}_\mathrm{O}$. Past the outer horizon, we have another timelike region. Meanwhile, the ergoregion structure is $\mathrm{E}$ between $\mathcal{E}_0$ and the outer ergosurface $\mathcal{E}_\mathrm{O}$ $(r^\mathcal{E}_0 < r < r^\mathcal{E}_\mathrm{O})$, followed by $\mathrm{N}$ outside $\mathcal{E}_\mathrm{O}$.The cosmological ergosurface $\mathcal{E}_\mathrm{C}$ and horizon $\mathrm{H}_\mathrm{C}$ are then found further out. These demarcate transitions $\mathrm{N} \rightarrow \mathrm{E}$ and $\mathrm{T} \rightarrow \mathrm{S}^+$ respectively. 

Below this in figure~\ref{fig: general features}, we have the extremal ergosurface cosmological limit $\mathcal{E}_{\mathrm{ex(c)}}$ where the outer $\mathcal{E}_\mathrm{O}$ and cosmological $\mathcal{E}_\mathrm{C}$ ergosurfaces coincide $r^\mathcal{E}_\mathrm{O} = r^\mathcal{E}_\mathrm{C}$. The causal and ergoregion structure remains identical to Domain IV. The existence of the $\mathrm{N}$ region between $\mathcal{E}_{\mathrm{O}}$ and $\mathcal{E}_{\mathrm{C}}$ is due to reasoning akin to the existence of the $\mathrm{S}^-$ region between the inner and outer horizons in the extremal spin limit. This limit then serves as the transition point on the map, wherein the ergoregion structure becomes just $\mathrm{E}$ for all $r > 0$ for $\gamma < \gamma^{\mathcal{E}}_{\mathrm{ex(c)}}$. This limit lies on a local minimum of the ergosurface curve. Given $\kappa = 0$ and~\eqref{eq: ergo extremal gamma}, we find $\gamma^{\mathcal{E}}_{\mathrm{ex(c)}} = -\frac{1}{3}$.

Domain V shares the causal structure of Domain IV, but with an ergoregion structure that is just $\mathrm{E}$.

The extremal horizon cosmological limit $\mathrm{H}_{\mathrm{ex(c)}}$ represents the lower $\gamma$ limit of black hole spacetimes, as the outer black hole event horizon $\mathrm{H}_{\mathrm{O}}$ coincides with the cosmological horizon $\mathrm{H}_{\mathrm{C}}$ here. This limit is found at a local minimum of the horizon curve.

Finally, in Domain VI, the inner horizon $\mathrm{H}_{\mathrm{I}}$ is now the sole remaining horizon. Its nature is different here, however. While it served as a Cauchy horizon in the black hole spacetimes, it now operates as the boundary between a $\mathrm{T}$ region and an $\mathrm{S}^+$ region. This now makes it a cosmological horizon. Since, as we discussed in subsection~\ref{subsection: extremal hor}, Cauchy and cosmological horizons both arise from \textit{repulsion}, this is unsurprising. We thus have naked singularity spacetimes in this domain.

\subsection{Role of $\gamma$}\label{subsection: role of gamma}
We see that the dependence of the CG Kerr solution ~\eqref{eq:CG Kerr} on $\gamma$, as seen in figure \ref{fig:Turner a = 0.25}, is rather complicated. We thus first turn to the CG Schwarzschild case~\eqref{eq:CG Schwarzschild}, where this dependence on $\gamma$ is more straightforward by considering that the relation between the lapse function $B(r)$~\eqref{eq:CG B} and the potential $\Phi(r)$ is $B(r) = 1 + 2\Phi(r)$. We then have an \textit{effective force} $\mathbf{f} = -\mathbf{\nabla\Phi}$, so that for large $r$, the linear $\gamma r$ term in the potential is attractive when $\gamma > 0$ and repulsive when $\gamma < 0$. Thus, in the $\kappa = 0$ case of the CG Schwarzschild solution, cosmological horizons are found for $\gamma < 0$, while none are found for $\gamma > 0$~\cite{Turner_2020}.

We also note that compared to the GR Schwarzschild case, the $-1/r$ term, dominating small $r$ behavior, in the CG Schwarzschild potential is modified by a factor of $(2-3\gamma)$ to become $-(2-3\gamma)/r$. This then switches sign to become positive when $\gamma > \frac{2}{3}$. However, as this becomes relevant at low $r$, the repulsive effect of this generates an inner Cauchy horizon instead of a cosmological horizon. Thus, even as the attractive nature of the $\gamma r$ term for $\gamma > 0$ dominates the large $r$ behavior, when $\kappa = 0$, an inner  Cauchy horizon $\mathrm{H}_{\mathrm{I}}$  may still be present behind an outer event horizon $\mathrm{H}_{\mathrm{O}}$. This is akin to the presence of the inner Cauchy horizon in GR Reissner-Nordstrom-AdS and GR Kerr-AdS, from the repulsive effect of charge and spin respectively, despite an attractive curvature $(\Lambda < 0)$ dominating the large $r$ behavior. 

Dependence on $\gamma$ is less straightforward when we seek to solve $\Delta^\mathrm{H} = 0$~\eqref{eq:rescaled} in the CG Kerr case. As we can see in ~\eqref{eq:rescaled}, no terms depend on merely the sign of $\gamma$ itself, but on more complicated combinations of it. For instance, the linear term in this quartic is proportional to $-(2-3\gamma)$. The effect of this term then depends on the sign of $(2-3\gamma)$ rather than just $\gamma$.

Furthermore, the lapse function $B(r)$~\eqref{eq:CG B} in the CG Schwarzschild case has $\kappa$ as the sole coefficient of the de Sitter quadratic term. Meanwhile, for CG Kerr, the coefficient $k$~\eqref{eq:auxiliary} of the now quartic de Sitter term in $\Delta^\mathrm{H}$ is composed of both $\kappa$ and combinations of $\gamma$.

To clarify this, we consider the case of $\kappa = 0$, where $k = k_0$~\eqref{eq: k0}. As all other factors are squared, the sign of $k_0$ is entirely determined by the sign of $(1- \gamma)$. Since this is always positive for $\gamma < \gamma_{max} = 1$, $k_0 > 0$ provides a repulsive effect akin to $\Lambda > 0$ in GR Kerr-dS. This thus allows cosmological horizons and $S^+$ regions to be found from Domains II downwards in figure~\ref{fig:Turner a = 0.25}. Therefore, unlike the CG Schwarzschild case~\cite{Turner_2020}, cosmological horizons may exist for $\gamma > 0$ when $\kappa = 0$. In fact, due to the $(2- 3\gamma)^2$ in the denominator, $k_0$ becomes very large as we approach $\gamma = \frac{2}{3}$, explaining the cosmological horizon $\mathrm{H}_\mathrm{C}$ curves emanating from $(r, \gamma) = (0,\frac{2}{3})$ in Domains II and III.

The same analysis presented here for horizons generally applies to the ergosurfaces and ergoregions governed by $\Delta^\mathcal{E}$~\eqref{eq:rescaled ergo} as well. When $\kappa \neq 0$, the interplay of $k_0$ and $\kappa$ determines the overall sign of $k$, which we discuss in the next section.

\section{Variation of $\kappa$}\label{section:kappa}

\begin{figure}[tb]
\centering
\includegraphics[width=1.05\linewidth]{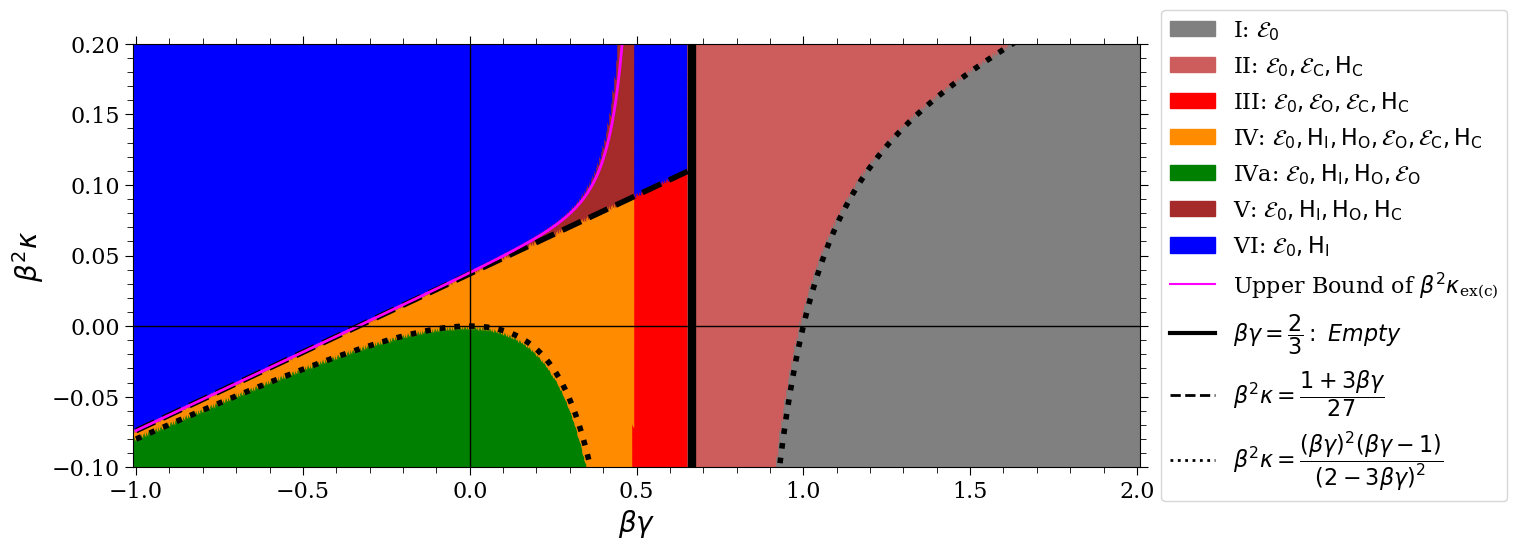}
\caption{Map of the parameter space of $\beta^2 \kappa$ against $\beta \gamma$, for spin $a/\beta = 0.25$, showing the relevant spacetime domains defined in tables~\ref{tab:domains} and~\ref{tab:domains neg}.}

\label{fig:Turner a = 0.25}
\end{figure}

\begin{figure}
    \centering
    \begin{subfigure}[b]{0.445\textwidth}
        \includegraphics[width=\textwidth]{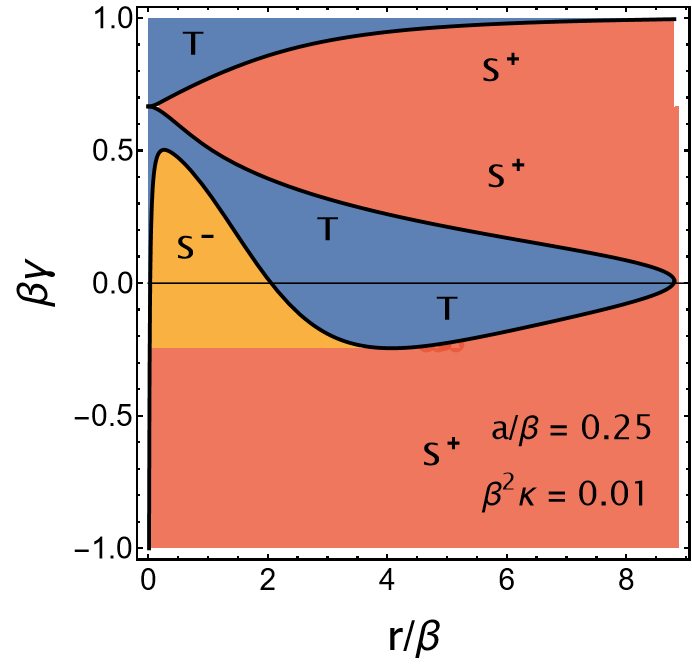}
        \caption[]%
        {Causal structure}    
        \label{fig: general causal2}
    \end{subfigure}
    \hfill
    \begin{subfigure}[b]{0.445\textwidth}  
        \includegraphics[width=\textwidth]{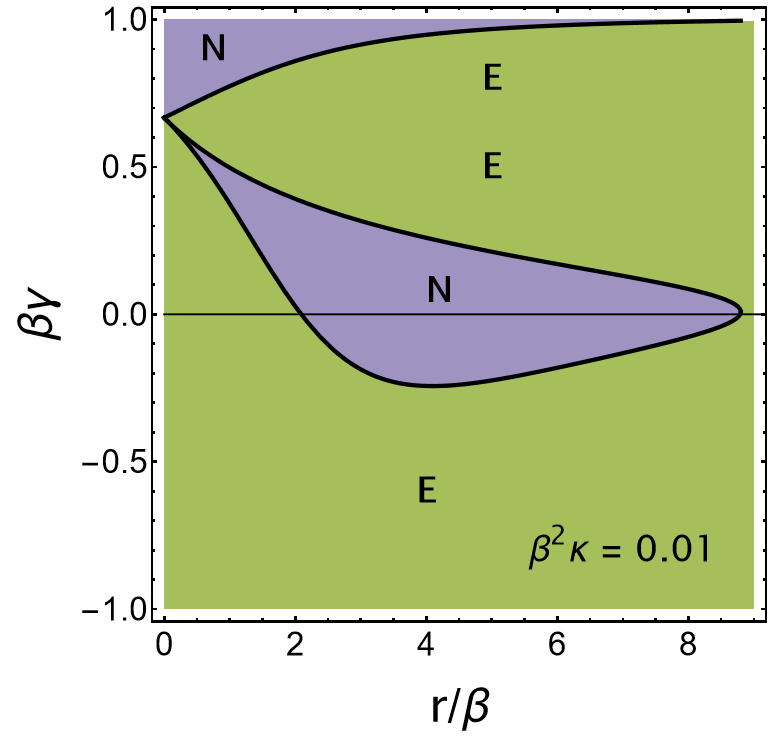}
        \caption[]%
        {Ergoregion structure}    
        \label{fig: general ergo2}
    \end{subfigure}
    \hfill
    \begin{subfigure}[b]{0.7\textwidth}  
        \includegraphics[width=\textwidth]{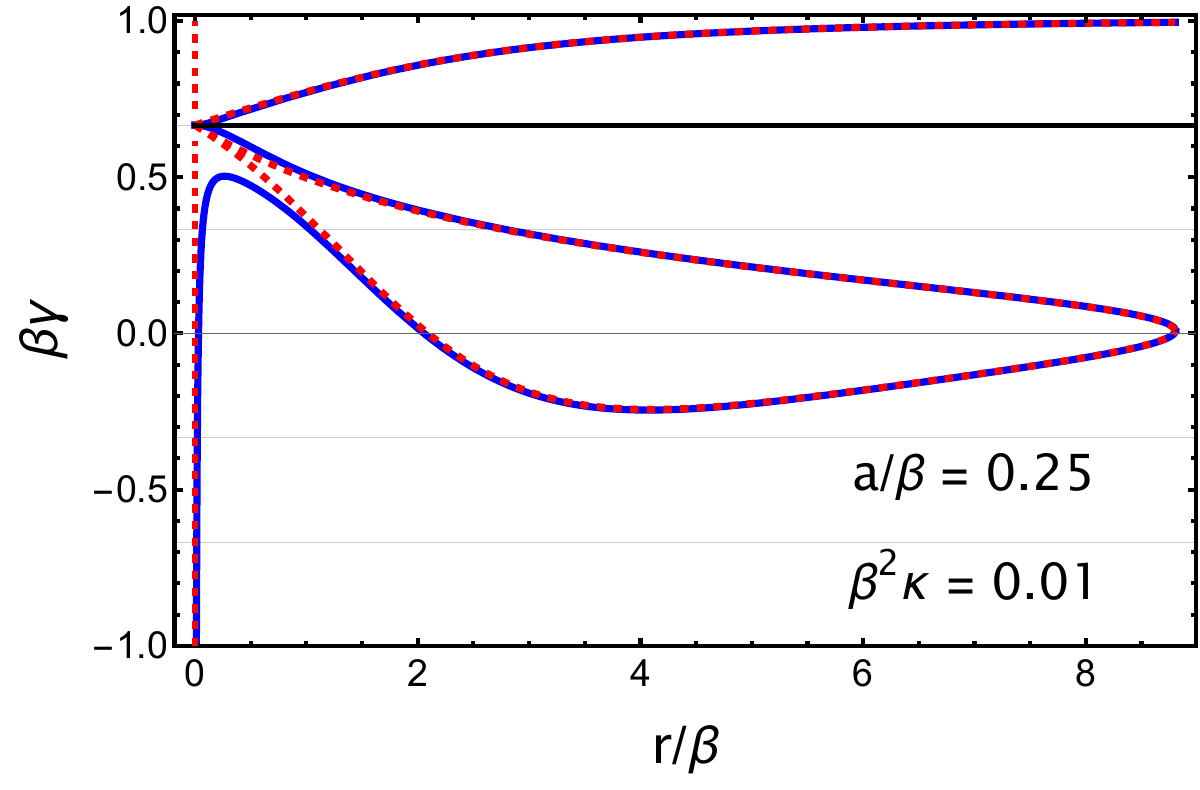}
        \caption[]%
        {Horizons (blue) and ergosurfaces (dashed red)}    
        \label{fig: kappa plus var}
    \end{subfigure}

    \caption[]
    {Causal structure, ergoregion structure, and plot of horizons (blue) and ergosurfaces (dashed red) of the CG Kerr spacetime for $(\beta^2\kappa, a/\beta) = (0.01, 0.25)$. The solid black line represents the \textit{Empty} case at $\beta \gamma = \frac{2}{3}$.} 
    \label{fig:kappa plus}
\end{figure}

\begin{figure}
    \centering
    \begin{subfigure}[b]{0.445\textwidth}
        \includegraphics[width=\textwidth]{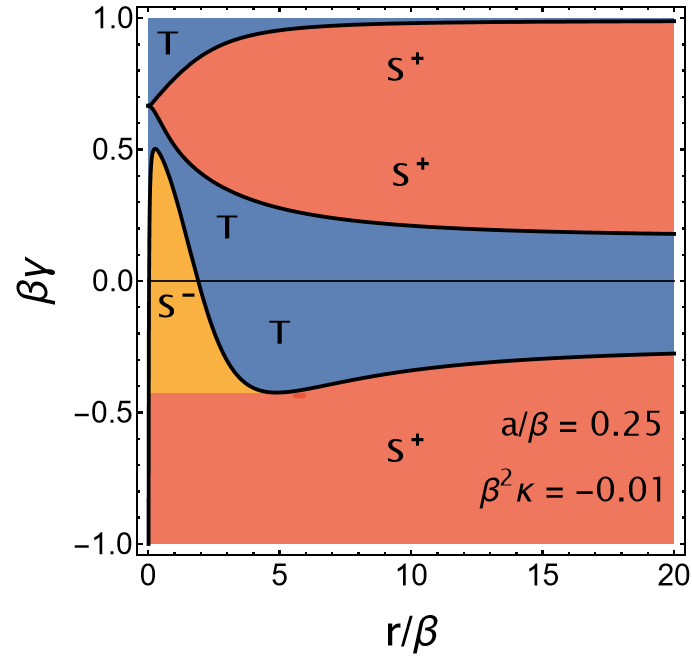}
        \caption[]%
        {Causal structure}    
        \label{fig: general causal3}
    \end{subfigure}
    \hfill
    \begin{subfigure}[b]{0.445\textwidth}  
        \includegraphics[width=\textwidth]{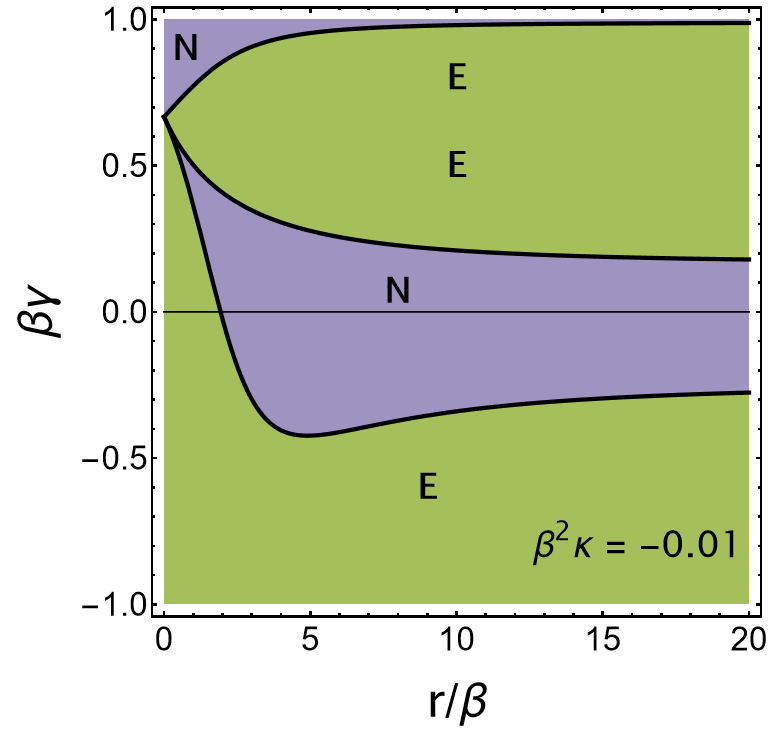}
        \caption[]%
        {Ergoregion structure}    
        \label{fig: general ergo3}
    \end{subfigure}
    \hfill
    \begin{subfigure}[b]{0.7\textwidth}  
        \includegraphics[width=\textwidth]{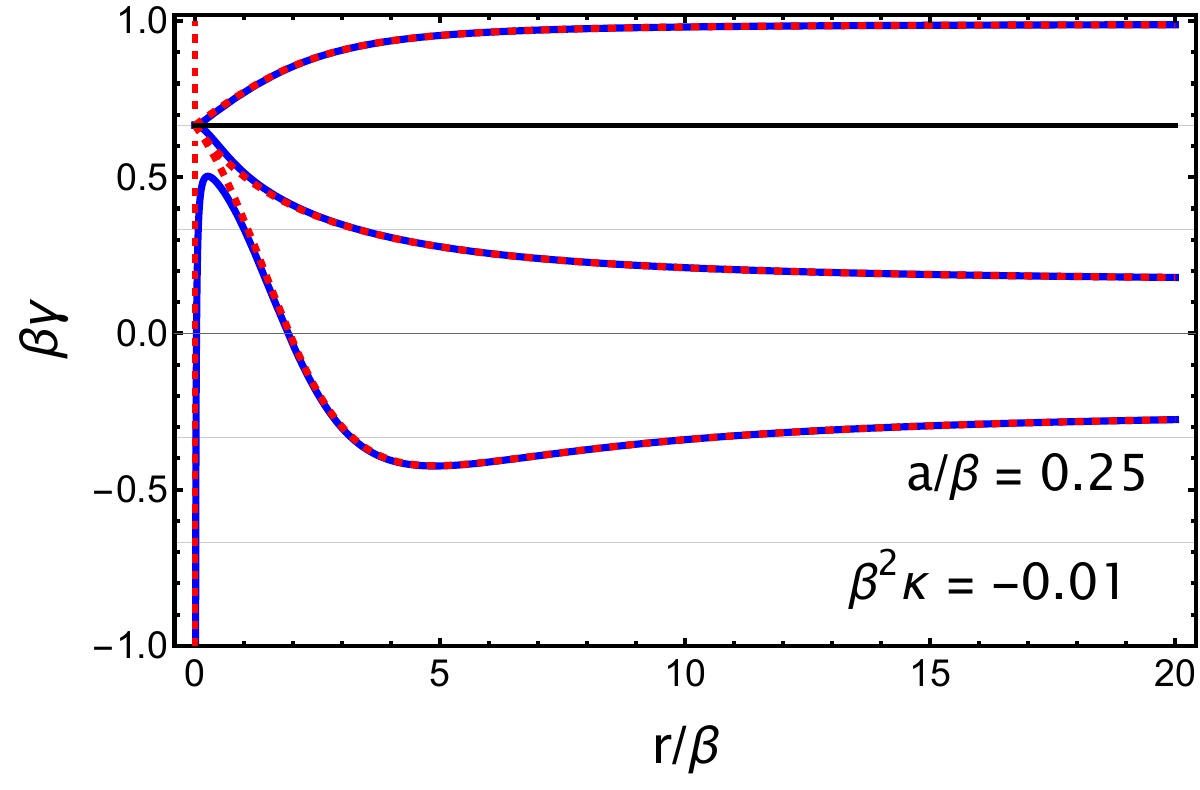}
        \caption[]%
        {Horizons (blue) and ergosurfaces (dashed red)}    
        \label{fig: kappa neg var}
    \end{subfigure}

    \caption[]
    {Causal structure, ergoregion structure, and plot of horizons (blue) and ergosurfaces (dashed red) of the CG Kerr spacetime for $(\beta^2\kappa, a/\beta) = (-0.01, 0.25)$. The solid black line represents the \textit{Empty} case at $\beta \gamma = \frac{2}{3}$.} 
    \label{fig:kappa minus}
\end{figure}

Proceeding now to the effects of varying $\kappa$, we present a map of spacetimes in the $\kappa$ vs $\gamma$ parameter space for $a = 0.25$ in figure~\ref{fig:Turner a = 0.25}. Here, we have color-coded each of the respective domains, and indicated some useful boundaries.

Additionally, we append the suffix ``a" to the domain name to indicate when the cosmological features ($\mathrm{H}_\mathrm{C})$ and $(\mathcal{E}_\mathrm{C}$) and regions ($\mathrm{H}_\mathrm{C}$ and $\mathrm{S}^+$) are no longer present as they have been pushed out \textit{beyond} $r = + \infty$, leaving them with an Anti-de Sitter  background. For instance, Domain IVa may be thought of as Domain IV but without the cosmological ergosurface $\mathcal{E}_\mathrm{C}$ and horizon $\mathrm{H}_\mathrm{C}$, and the corresponding cosmological ergoregion $\mathrm{E}$ and cosmological spacelike region $\mathrm{S}^+$. This thus gives Domain IVa the structure of black hole spacetimes of GR Kerr-AdS. We present the additional domains appearing with $\kappa < 0$ in table~\ref{tab:domains neg}.

We have four relevant domain boundaries in figure~\ref{fig:Turner a = 0.25}. Firstly, we have $\gamma = \frac{2}{3}$ denoting the \textit{Empty} spacetime discussed earlier. 

Secondly, we have the dashed diagonal line separating Domain IV from Domain V and Domain III from VI. Looking at table~\ref{tab:domains}, we see that the transition between IV and V, when the extremal ergosurface cosmological limit $\mathcal{E}_{\mathrm{ex(c)}}$ is crossed, has the loss of the outer $\mathcal{E}_{\mathrm{O}}$ and cosmological $\mathcal{E}_{\mathrm{C}}$ ergosurfaces. We see the same loss of these ergosurfaces when going from III to VI. While VI has the horizon labelled as $\mathrm{H}_\mathrm{I}$, we recall that it serves as a cosmological horizon here. This line is thus given by~\eqref{eq: UPS}, from the vanishing of $D^{\mathcal{E}}$~\eqref{eq:ergodisc}. The condition in~\eqref{eq:gamma ex restric} restricts this line to $\gamma < \frac{2}{3}$, as can be seen.

The third domain boundary, given by the dotted curves, marks the transitions I $\rightarrow$ II for $\gamma > \frac{2}{3}$, and IVa $\rightarrow$ IV when $\gamma < \frac{2}{3}$. These transitions are characterized by the addition of a cosmological ergosurface $\mathcal{E}_{\mathrm{C}}$ and horizon $\mathrm{H}_{\mathrm{C}}$. This derives from the changing of the sign of $k$ in the quartic de Sitter term in both $\Delta^{\mathcal{E}}$~\eqref{eq:rescaled ergo} and $\Delta^{\mathrm{H}}$~\eqref{eq:rescaled}. Solving then for $k = 0$~\eqref{eq:auxiliary}, we have the equation for the dotted curves
\begin{equation}\label{eq: SPS}
    \kappa = -k_0 = \frac{\gamma^2(\gamma - 1)}{(2-3\gamma)^2}.
\end{equation}

The switch from $k < 0$ to $k > 0$ when $\kappa > - k_0$ makes the quartic term in both  $\Delta^{\mathcal{E}}$~\eqref{eq:rescaled ergo} and $\Delta^{\mathrm{H}}$~\eqref{eq:rescaled} repulsive, generating the corresponding cosmological ergosurface $\mathcal{E}_{\mathrm{C}}$ and horizon $\mathrm{H}_{\mathrm{C}}$.

For $\gamma > \frac{2}{3}$, this boundary~\eqref{eq: SPS} is, in fact, given by $\gamma_{\mathrm{max}}$, discussed in the previous section. From this, we see that increasing $\kappa > 0$ acts to increase $\gamma_{\mathrm{max}}$, while making $\kappa$ more negative decreases $\gamma_{\mathrm{max}}$.

Turning now to $\gamma < \frac{2}{3}$, we see that underneath the curve defined by~\eqref{eq: SPS} we have Domain IVa, while above we have Domain IV. We then understand that spacetimes below the dotted curves defined by~\eqref{eq: SPS} have Anti-de Sitter backgrounds, while those above have de Sitter backgrounds. 

The final boundary we have denoted in figure~\ref{fig:Turner a = 0.25} relates to the extremal horizon cosmological limit $\mathrm{H}_{\mathrm{ex(c)}}$, as it marks a transition V $\rightarrow$ VI. From the discussion in subsection~\ref{subsection: extremal hor}, we know that the relevant boundary would be found by a solution to $D^{\mathrm{H}} = 0$~\eqref{eq:hordisc}. From matching the solutions to what we see in Figure~\ref{fig:Turner a = 0.25}, we find that the upper $\kappa$ limit of the Domain V region is given by 
\begin{eqnarray}\label{eq: upper kappa}
 \fl\kappa_{\mathrm{ex(c)}} = \frac{1}{512} \left(-\frac{27 (2-3 \gamma )^4}{a^6}+\frac{144 (2-3 \gamma
   )^2}{a^4}-\frac{128}{a^2} + 512k_0\right) \nonumber\\
   +\frac{1}{512} \ \sqrt{\frac{(2-3 \gamma )^2 \left(9 (2-3 \gamma )^2-32
   a^2\right)^3}{a^{12}}}.
\end{eqnarray}
This is of course limited to $\gamma < \frac{2}{3}$ from the restriction in~\eqref{eq:gamma ex restric}. While this represents the upper edge of the Domain V region in the map, we failed to find a similarly useful expression for the right edge of Domain V. 

\begin{table}

\caption{\label{tab:domains neg} Additional domains, not described by table~\ref{tab:domains}, that are relevant for $\kappa \leq 0$. Here, $\gamma^{\mathrm{H}}_{\mathrm{ex(s)}}$ is the value of $\gamma$ at the extremal spin limit. The spacetimes described by these domains all have an Anti-de Sitter background.
}

\begin{tabular}{@{}lllll}
\br
\rm Label &Domain & Features Present & Causal Structure & Ergoregions\\
\mr
IIIa&$\gamma >\gamma^{\mathrm{H}}_{\mathrm{ex(s)}}$ & $\mathcal{E}_0 , \mathcal{E}_\mathrm{O}$ & $\mathrm{T} $ & $ \mathrm{E} \rightarrow \mathrm{N}$\\

$\mathrm{H}_{\mathrm{ex(s)}}$a &$\gamma =\gamma^{\mathrm{H}}_{\mathrm{ex(s)}}$ &$\mathcal{E}_0 , \mathrm{H}_{\mathrm{I}} = \mathrm{H}_{\mathrm{O}} ,  \mathcal{E}_{\mathrm{O}}$ & $\mathrm{T} \rightarrow \mathrm{S}^- \rightarrow \mathrm{T}$ & $\mathrm{E} \rightarrow \mathrm{N}$\\

IVa &$\gamma <\gamma^{\mathrm{H}}_{\mathrm{ex(s)}}$ &$\mathcal{E}_0 , \mathrm{H}_{\mathrm{I}} , \mathrm{H}_{\mathrm{O}} , \mathcal{E}_{\mathrm{O}}$ & $\mathrm{T} \rightarrow \mathrm{S}^- \rightarrow \mathrm{T}$ & $\mathrm{E} \rightarrow \mathrm{N}$\\

\br
\end{tabular}

\end{table}

We also failed to find a closed form expression for the extremal spin limit $\mathrm{H_{\mathrm{ex(s)}}}$, defining the transition between the Domain III and IV spacetimes in figure~\ref{fig:Turner a = 0.25}. We therefore solved for these using the procedure outlined in subsection~\ref{subsection: extremal hor}. 

Having described the indicated domains, we may now analyze the effect of increasing $\kappa$ on the spacetime structure. Turning to figure~\ref{fig:kappa plus}, with $(\kappa, a) = (0.01, 0.25)$, we see that the increase of $\kappa$ leads to the cosmological ergosurfaces $\mathcal{E}_{\mathrm{C}}$ and horizons $\mathrm{H}_{\mathrm{C}}$ being pulled in to smaller $r$. In comparison to the $\kappa = 0$ case in figure~\ref{fig:General Structure}, we see that even the $\gamma = 0$ case has the cosmological horizon $\mathrm{H}_\mathrm{C}$ and ergosurface $\mathcal{E}_\mathrm{C}$ at finite $r$. 

Solving for $\gamma_{\mathrm{max}}$ from $k = 0$~\eqref{eq:auxiliary}, we see that it increases above $ \gamma_{\mathrm{max}} = 1$ when we bring up $\kappa > 0$. This is as $k$~\eqref{eq:auxiliary} will then switch sign at a higher value of $\gamma$.

Additionally, we see that the $\gamma$ value at the extremal ergosurface cosmological limit $\gamma^{\mathcal{E}}_{\mathrm{ex(c)}}$ has increased, as can be confirmed from~\eqref{eq: ergo extremal gamma}. While we do not have a similar closed form expression for the $\gamma$ value at the extremal horizon cosmological limit, we also observe this increasing when comparing figures~\ref{fig:General Structure} $(\kappa = 0)$ and~\ref{fig:kappa plus} $(\kappa = 0.01)$. 

When we now turn to $\kappa < 0$, domains containing the cosmological features $\mathrm{H}_\mathrm{C}$ and $\mathcal{E}_\mathrm{C}$ are still present. For instance, we still have some Domain IV spacetimes in figure~\ref{fig:Turner a = 0.25}, outside the region defining Domain IVa. This is as the quartic de Sitter term may still remain repulsive as long as $k > 0$ corresponding to $\kappa > -k_0$, with the transition, of course, determined by~\eqref{eq: SPS}. In general, however, making $\kappa$ more negative acts to push $\mathrm{H}_\mathrm{C}$ and $\mathcal{E}_\mathrm{C}$ further out, as can be seen in figure~\ref{fig:kappa minus}. 

This then also manifests in decreasing the value of $\gamma_{\mathrm{max}}$ for more negative $\kappa$. We see a similar lowering of the  $\gamma$ values at the extremal ergosurface cosmological limit $\gamma^{\mathcal{E}}_{\mathrm{ex(c)}}$~\eqref{eq: ergo extremal gamma} and the extremal horizon cosmological limit when comparing figures~\ref{fig:General Structure} and~\ref{fig:kappa minus}.

\section{Dependence on the spin $a$}\label{section:spin}

\begin{figure}[htp]
\centering 

\subfloat[$\dfrac{a}{\beta} = 0.01$]{%
  \includegraphics[clip,width=1.05\linewidth]{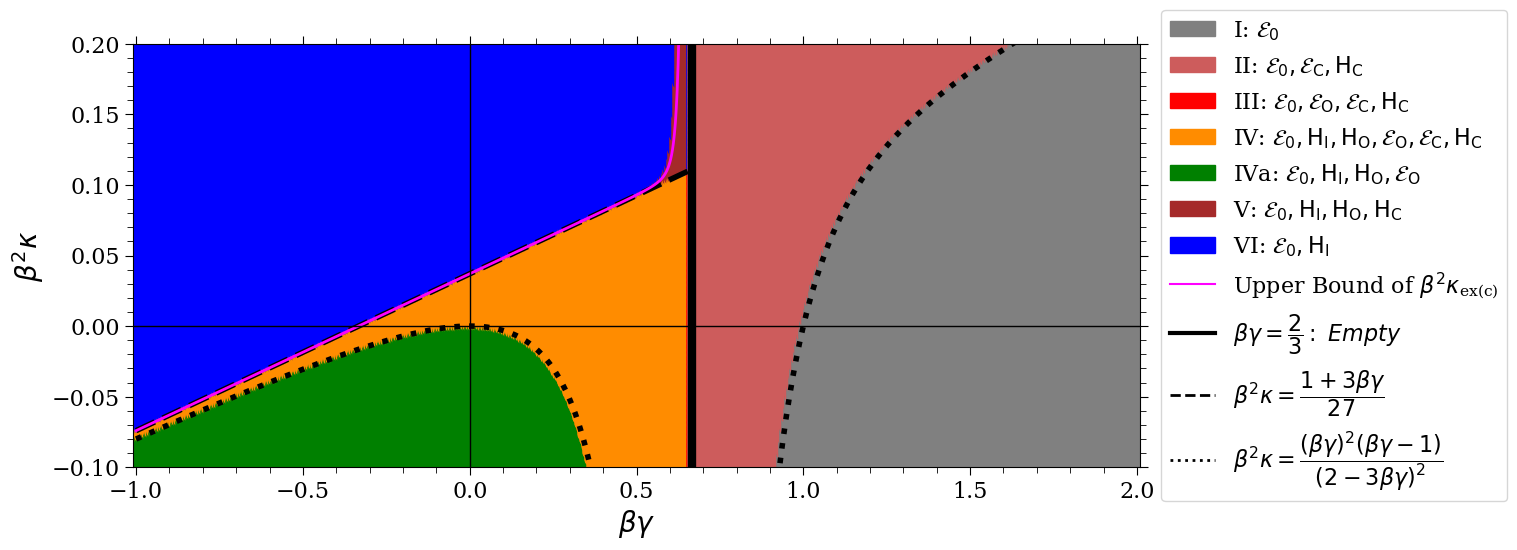}\label{fig:a=0.01}
}

\subfloat[$\dfrac{a}{\beta} = 1$]{%
  \includegraphics[clip,width=1.05\linewidth]{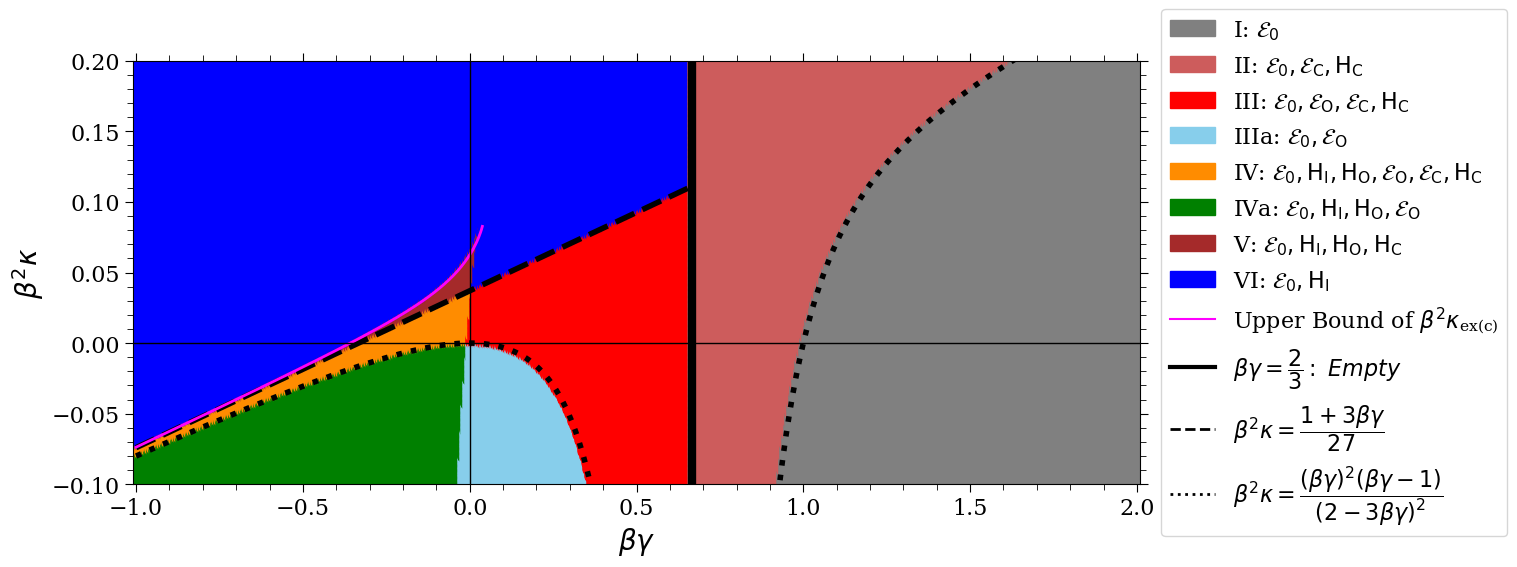}\label{fig:a=1}
}

\subfloat[$\dfrac{a}{\beta} = 1.5$]{%
  \includegraphics[clip,width=1.05\linewidth]{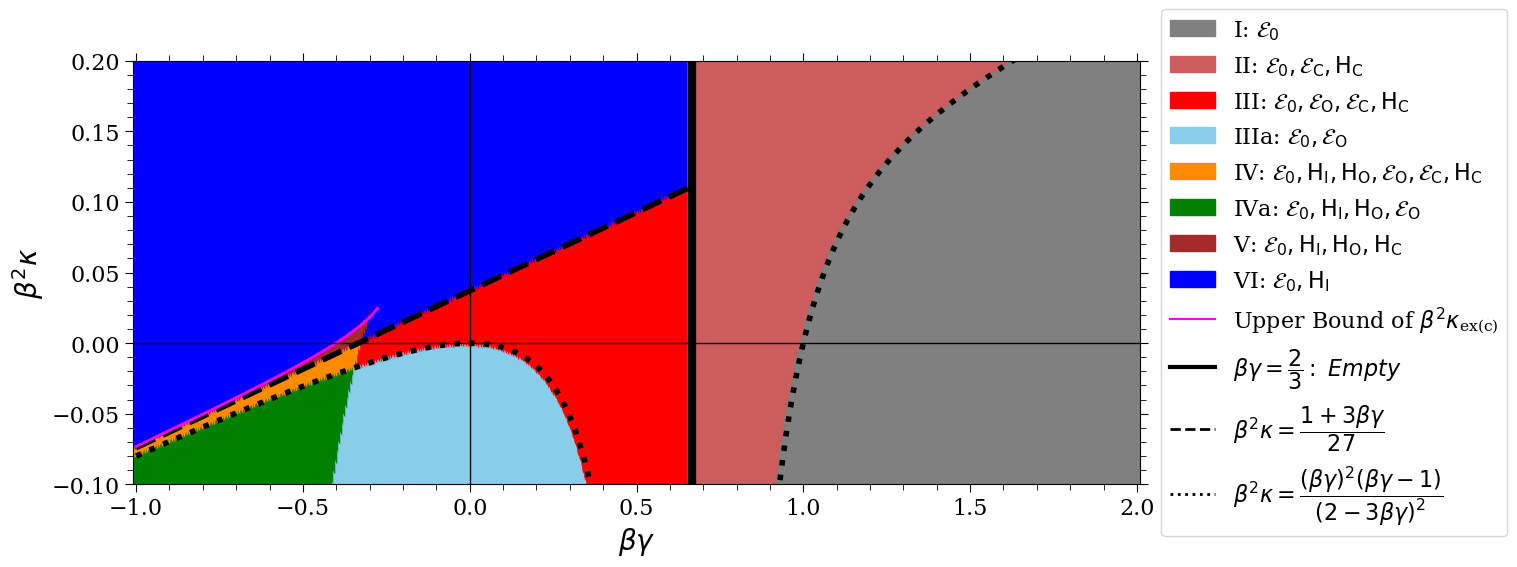}\label{fig:a=1.5}
}

\caption{Map of the parameter space of $\beta^2 \kappa$ against $\beta \gamma$, for varying spin $a/\beta$, showing the relevant spacetime domains defined in tables~\ref{tab:domains} and~\ref{tab:domains neg}. }

\label{fig:Turner spin}
\end{figure}

\begin{figure}[tb]
\centering
\includegraphics[width=1.05\linewidth]{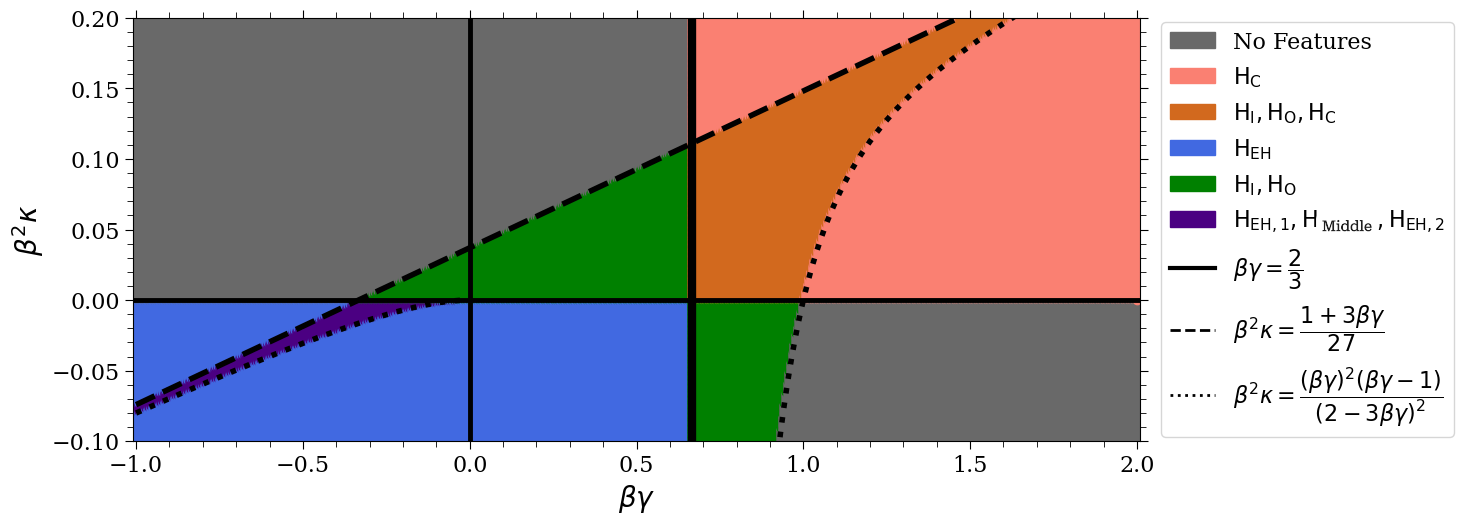}
\caption{Map of the parameter space of $\beta^2 \kappa$ against $\beta \gamma$, for the CG Schwarzschild case of spin $a/\beta = 0$. Features present in different color-coded regions are described in the side panel.}

\label{fig:Turner CG Schwarzschild}
\end{figure}

Figure~\ref{fig:Turner spin} demonstrates the effect of varying spin $a$ on the spacetimes in the parameter space of $\gamma$ and $\kappa$. Obviously, as $\Delta^{\mathcal{E}}$~\eqref{eq:rescaled ergo} has no dependence on $a$, we see no changes in the presence and ergosurfaces as we increase spin $a$.

One may first notice that the domains present for $\gamma > \frac{2}{3}$ are unaffected as we increase the spin from $a = 0.01$ to $a = 1.5$. As the demarcation between Domain I and Domain II in this region of parameter space is entirely determined by~\eqref{eq: SPS}, which is not spin $a$ dependent, this is not surprising.

Turning to $\gamma < \frac{2}{3}$, we see clear changes as we increase the spin. Spacetimes covered by Domains IV, IVa, and V move down and to the left, or to lower $\gamma$ and $\kappa$ as $a$ is increased. It may be recalled that the transition V $\rightarrow$ VI is given by the extremal horizon cosmological limit $\mathrm{H}_{\mathrm{ex(c)}}$. We then see the Domain V region moving further down along the dashed line given by~\eqref{eq: UPS} as spin $a$ is increased. Meanwhile the transition III $\rightarrow$ IV is of course described by the extremal spin limit $\mathrm{H}_{\mathrm{ex(s)}}$.

Domain IIIa is of course is a horizonless spacetime with only the ergosurfaces $\mathcal{E}_0$ and $\mathcal{E}_\mathrm{O}$ present. This thus has the causal structure of simply $\mathrm{T}$ for $r > 0$. Its ergoregion structure meanwhile is identical to Domain IVa as $\mathrm{E} \rightarrow \mathrm{N}$.

The transition IVa $\rightarrow$ IIIa is then also described by the extremal spin limit, with the merging and then annihilation of the inner Cauchy horizon $\mathrm{H}_{\mathrm{I}}$ and outer event horizon $\mathrm{H}_{\mathrm{O}}$. However, as IIIa and IVa have AdS backgrounds, we note that the extremal spin limit spacetime separating the two would also have such an AdS background, so we refer to it as $\mathrm{H}_{\mathrm{ex(s)}}$a in table~\ref{tab:domains neg}.

As mentioned, we see the Domain IV and IVa regions moving to lower and lower $\gamma$ and $\kappa$ as spin $a$ is increased. What is clear from the $a = 1.5$ in figure~\ref{fig:a=1.5} is that black hole spacetimes described by these two regions still exist past the GR Kerr extremal spin limit of $a = 1$. 

Conversely, when looking at figure~\ref{fig:Turner a = 0.25} $(a = 0.25)$ and~\ref{fig:a=1} $(a = 1)$, many Domain IV or IVa spacetimes have already been taken over by Domain III and IIIa spacetimes respectively. We may thus say that Kerr black holes in conformal gravity may exist above and below the GR Kerr extremal spin limit of $a = 1$, depending on the values of $\gamma$ and $\kappa$. This is similar to how GR Kerr-dS spacetimes have  extremal spin values above $a = 1$~\cite{kerrdSextremal}, while GR Kerr-AdS spacetimes have values below $a = 1$.

Similarly, the extremal horizon cosmological limit controls the existence of Domain V black holes. In GR Kerr-dS spacetimes, this limit is of course dependent on both the cosmological constant $\Lambda$ and the spin $a$. However, there is an upper bound for the cosmological constant for this limit in GR Kerr-dS given by $\Lambda_{\mathrm{max}} \approx 0.1778$~\cite{kerrdSextremal}. In CG Kerr, if we take $\gamma = 0$, then $k = \kappa = \Lambda/3$, and this maximum value would translate to $\kappa_{\mathrm{max}} \approx 0.059267$. 

When we once again include $\gamma \neq 0$, both figures~\ref{fig:Turner a = 0.25} and~\ref{fig:Turner spin} demonstrate that that the additional dependence on the $\gamma$ parameter in CG Kerr spacetimes means that black holes may exist for $\kappa$ values above this $\kappa_{\mathrm{max}} \approx 0.059267$ limit of GR Kerr-dS.

\subsection{Comparison to CG Schwarzschild}

We now compare the CG Kerr spacetimes we have thus far discussed to the $a = 0$ case of CG Schwarzchild by similarly presenting a map of its spacetimes in the $\kappa$ vs $\gamma$ parameter space in figure~\ref{fig:Turner CG Schwarzschild}. This reproduces the horizon structure of the map in~\cite{Turner_2020}. As our work focuses on the CG Kerr solution, we do not go into fully detailing the CG Schwarzschild spacetimes here. We direct readers interested in a more comprehensive discussion of such to~\cite{ Turner_2020, villanueva}.

It must again be stressed that the CG Kerr metric~\eqref{eq:CG Kerr} does not trivially reduce to the CG Schwarzschild metric~\eqref{eq:CG Schwarzschild} when we take $a \rightarrow 0$. A conformal and coordinate transformation is necessary to bring the CG Kerr metric~\eqref{eq:CG Kerr} with $a = 0$ to the required form of the CG Schwarzschild metric~\eqref{eq:CG Schwarzschild}. Such transformations have not explicitly been derived for the CG Kerr metric in Boyer-Lindquist coordinates, so readers interested in seeing the full transformations may look at~\ref{section: Appendix conformal}.

An immediate difference between CG Schwarzschild and Kerr is that the ergosurfaces and horizons are one and the same in CG Schwarzschild, while they are distinct features in CG Kerr. While any \textit{repulsive effect} in CG Schwarzschild must be due to $\gamma$ and/or $\kappa$, the presence of spin $a$ in CG Kerr means that at least some repulsion is ubiquitous in CG Kerr spacetimes.

Looking at figure~\ref{fig:Turner CG Schwarzschild}, we see three boundaries also found in CG Kerr. Firstly, the line $\gamma = \frac{2}{3}$ is once again notable here. While in CG Kerr, we have the \textit{Empty} case that we believe may be a degeneracy, the solution is unproblematic in CG Schwarzschild, where crossing from $\gamma < \frac{2}{3}$ to $\gamma > \frac{2}{3}$ sees the addition of a \textit{repulsive horizon}. This is a cosmological horizon $\mathrm{H}_{\mathrm{C}}$ for $\kappa > 0$ and an inner Cauchy horizon $\mathrm{H}_{\mathrm{I}}$ for $\kappa < 0$.

Next, we see the familiar dashed diagional line given by~\eqref{eq: UPS}. While in CG Kerr, this boundary is confined to $\gamma < \frac{2}{3}$ due to~\eqref{eq:gamma ex restric}, no such restriction exists for CG Schwarzschild. This boundary thus extends past $\gamma = \frac{2}{3}$ in figure~\ref{fig:Turner CG Schwarzschild}. Crossing from below this diagonal line to above it results in the coalescence of the Cauchy horizon of a black hole with its event horizon, leading them to mutually annihilate. Meanwhile, in CG Kerr, for $\gamma < \frac{2}{3}$,~\eqref{eq: UPS} marks the annihilation of an outer ergosurface $\mathcal{E}_{\mathrm{O}}$ with a cosmological ergosurface $\mathcal{E}_{\mathrm{C}}$ . 

Thirdly, we once again see the dotted curves given by~\eqref{eq: SPS}. Unlike the CG Kerr case however, as may be seen in figures~\ref{fig:Turner a = 0.25} and~\ref{fig:Turner spin}, this dotted line is not found for $0 < \gamma < \frac{2}{3}$. As discussed in~\cite{Turner_2020}, the curve given by~\eqref{eq: SPS} for this range of $\gamma$ concerns a transition for a horizon found at $r < 0$. Since we do not concern ourselves with $r < 0$, it is not entirely surprising that we do not see a change in our CG Kerr maps here.

Furthermore, while in the CG Kerr maps in figures~\ref{fig:Turner a = 0.25} and~\ref{fig:Turner spin}, the spacetimes under the dotted curves all possess Anti-de Sitter backgrounds, this is not always true in the CG Schwarzschild map in figure~\ref{fig:Turner CG Schwarzschild}. For instance, the region under the dotted curve but above $\kappa = 0$ shows a cosmological horizon $\mathrm{H}_\mathrm{C}$ indicating a de Sitter background.

In the CG Schwarzschild parameter space, crossing from $\kappa < 0$ to $\kappa > 0$ sees the addition of a \textit{repulsive horizon}, either Cauchy or cosmological. This special transition when crossing $\kappa = 0$ is not present in CG Kerr spacetimes. 

We may attribute this to the aforementioned fact that the spin $a$ makes some repulsion ubiquitous, thus not making crossing to $\kappa > 0$ particularly noteworthy. 

Most importantly, as we have discussed extensively, while the sign of the quadratic de Sitter term in CG Schwarzschild~\eqref{eq:CG Schwarzschild} depends only on $\kappa$, the sign of the quartic de Sitter term in CG Kerr~\eqref{eq:CG Kerr} depends on the sign of $k$~\eqref{eq:auxiliary}. Thus, the background in CG Kerr is de Sitter when $ k > 0$, flat when $k = 0$, and Anti-de Sitter when $k < 0$.

While black hole spacetimes in CG Kerr always have at least two horizons, namely the inner Cauchy horizon $\mathrm{H}_\mathrm{I}$ and outer event horizon $\mathrm{H}_\mathrm{O}$, the medium blue regions in figure~\ref{fig:Turner CG Schwarzschild} show Schwarzschild black holes with just the event horizon. This difference is of course attributed to the repulsion from the spin $a$ in CG Kerr spacetimes being absent for CG Schwarzschild.

The black holes described by the violet domain in figure~\ref{fig:Turner CG Schwarzschild} are of note. These can be thought of as nested black holes. While we have an outer event horizon $\mathrm{H}_{\mathrm{EH,}2}$, a middle horizon $\mathrm{H}_{\mathrm{Middle}}$, there is another innermost event horizon $\mathrm{H}_{\mathrm{EH,}1}$. Thus, the causal structure of this domain is $\mathrm{S}^- \rightarrow \mathrm{T} \rightarrow \mathrm{S}^- \rightarrow \mathrm{T}$. While this intermediate horizon $\mathrm{H}_{\mathrm{Middle}}$ demarcates a transition $\mathrm{T} \rightarrow \mathrm{S}^-$ as $r$ is increased, and thus is another \textit{repulsive horizon}, we do not call it a Cauchy horizon because the $\mathrm{T}$ region it encloses does not possess the singularity. The absence of nested black holes in CG Kerr can again be ascribed to the repulsive spin $a$ dominating the small $r$ behavior, and thus preventing the existence of an innermost $\mathrm{S}^-$ region.

In general, then, the differences between the spacetimes present in the $\kappa$ vs. $\gamma$ parameter space of CG Schwarzschild and CG Kerr may be attributed to three things. Firstly, ergosurfaces and horizons are distinct in CG Kerr but are coincident in CG Schwarzschild. Secondly, we have the ubiquity of repulsion from the spin $a$ in CG Kerr. Finally, the causal structure of the background is determined by the sign of $k$~\eqref{eq:auxiliary} in CG Kerr, while it is only dependent on the sign of $\kappa$ in CG Schwarzschild.

\section{Conclusions and future prospects}\label{section:conclusions}

In this work, we elucidated the structure of spacetimes arising in the $\gamma-\kappa-a$ parameter space of the conformal gravity Kerr metric, for positive mass $(\beta > 0)$ as evaluated on the equatorial plane $(\theta = \frac{\uppi}{2})$. We found thirteen distinct configurations of these spacetimes. We further noted that the causal and ergoregion structure of the black hole spacetimes of CG Kerr have analogues in the Kerr-de Sitter and Kerr-Anti-de Sitter black hole spacetimes of general relativity, depending keenly on the values of $\gamma, \kappa,$ and $a$. 

We also derived formulae for the extremal spin limit, extremal horizon cosmological limit, and extremal ergosurface cosmological limit. This allowed us to determine that due to the additional dependence on the $\gamma$ parameter, spacetime solutions of the conformal gravity Kerr metric may possess extremal limit values lying above or below those of the analogous general relativity solutions, depending of course on the specific combination of parameters.

In addition to this, we calculated the surface gravities and temperatures associated with Hawking radiation for the horizons in the CG Kerr spacetimes. We found that these quantities both vanish at the extremal horizon limits, as one would also see in general relativity.

Future work could look at expanding our classification of spacetimes to $\beta < 0$ and to the $r < 0$ region within the ring singularity. As has been done for the CG Schwarzschild metric~\cite{Turner_2020}, it may be interesting to see if an appropriate conformal transformation can render curvature scalars at the CG Kerr ring singularity regular.

Additionally, as in~\cite{Turner_2020}, it would be of interest to explore finding the equatorial photon circles within this metric, and how they may further subclassify the spacetimes we found here. Studying the efficiency of energy extraction via the Penrose process would also be worthwhile to consider for this metric compared to the corresponding metrics of GR.

As extremal spacetimes are of course of interest to aspects of string theory and quantum gravity, studying the near-horizon geometries at the extremal cases found here may be of much interest. We mentioned earlier that no corollary cosmological Nariai metric~\cite{NariaiOriginal, NariaiHolo} has been found in conformal gravity, so exploring whether the near-horizon geometry of the extremal horizon cosmological limit we found here may allow for an initial step at deriving such a solution.

\section*{Acknowledgments}
Parts of this work were adapted from an MSc Dissertation report submitted by Miguel Yulo Asuncion under the supervision of co-authors Dominik and Horne.

We also thank Philip D. Mannheim for some instructive insights into conformal gravity.

\appendix

\section{Conformal transformation in the zero spin limit}\label{section: Appendix conformal}\setcounter{section}{1}

We now present an appropriate conformal transformation to demonstrate that the zero spin limit $(a \rightarrow 0)$ of the CG Kerr metric~\eqref{eq:CG Kerr} is conformally equivalent to the CG Schwarzschild metric~\eqref{eq:CG Schwarzschild}. 

We note that the $g_{tt}$ component of the CG Kerr metric~\eqref{eq:CG Kerr}, where $r$ is the Boyer-Lindquist radial coordinate~\cite{boyerlindquist}, is given by 
\begin{equation}
    g_{tt} = - \left( 1 - \frac{2 \widetilde{M} r}{r^2 + a^2 \cos^2{\theta}} - k\left(r^2 - a^2 \cos^2{\theta}\right) \right).
\end{equation}
When we set $a = 0$, we find that this reduces to
\begin{equation}
   g_{tt} = -B(r) = -\left(1 - \frac{2 \widetilde{M}}{r} - kr^2 \right).
\end{equation}
It is also readily shown that the radial component of the metric likewise reduces to $g_{rr} = 1/B(r)$, and that the cross terms $g_{t\phi} = g_{\phi t}$ vanish. The angular terms also reduce to their spherically-symmetric forms  $g_{\theta \theta} = r^2$ and $g_{\phi \phi} = r^2 \sin^2 \theta$ as required.

We do not see a linear potential as we would expect from the CG Schwarzschild metric~\eqref{eq:CG Schwarzschild}. To bring this to a form that has the said linear potential, we now consider the conformal transformation used in~\cite{KeithRotation}. Here, we have a conformal factor given by
\begin{equation}
    \Omega \equiv \frac{R + A}{A},
\end{equation}
such that $\Omega^2 \ g_{\mu \nu} = \widetilde{g}_{\mu \nu}$, where $\widetilde{g}_{\mu \nu}$ is the metric we shall show to be equivalent to CG Schwarzschild ~\eqref{eq:CG Schwarzschild}, and $A$ is a scale length. We then have the transformation
\begin{equation}\label{eq:R def}
   r = \frac{R}{\Omega} = \frac{RA}{R + A}.
\end{equation}

Rewriting~\eqref{eq:CG Schwarzschild}, now with the radial coordinate as $R$, the lapse function of $\widetilde{g}_{\mu \nu}$ is 
\begin{equation}\label{eq:CG Schwarzschild Lapse R}
    \widetilde{B}(R) = 1-3 \beta \gamma-\frac{\beta(2-3 \beta \gamma)}{R}+\gamma R-\kappa R^2.
\end{equation}

We wish to test if indeed $\Omega^2 B(r) = \widetilde{B}(R)$. We see that
\begin{equation}
    \Omega^2 B(r) = \Omega^2 - \Omega^3 \ \frac{2\widetilde{M}}{R} - kR^2.
\end{equation}
Rearranging this, we have 
\begin{equation}\label{eq:Omega2BrA}
     \Omega^2 B(r) = \left(1-\frac{6 \widetilde{M}}{A}\right)-\frac{2 \widetilde{M}}{R}+\frac{1}{A}\left(2-\frac{6 \widetilde{M}}{A}\right) R-\left(-\frac{1}{A^2}+\frac{2 \widetilde{M}}{A^3}+k\right) R^2.
\end{equation}
Identifying this with~\eqref{eq:CG Schwarzschild Lapse R}, we have
\begin{equation}
\eqalign{m = \widetilde{M}, \cr
\gamma = \frac{1}{A}\left(2 - \frac{6\widetilde{M}}{A} \right), \cr
\kappa = -\frac{1}{A^2} + \frac{2\widetilde{M}}{A^3} + k,}
\end{equation}
and scale length
\begin{equation}
A=\frac{2 - 3\beta \gamma}{\gamma}.
\end{equation}
Using this, we also find that we recover the definition of $k$ from~\eqref{eq:auxiliary} as
\begin{equation}
    k =  \kappa + \frac{1}{A^2} - \frac{2\widetilde{M}}{A^3} = \kappa - \frac{2\widetilde{M}-A}{A^3}.
\end{equation}
Substituting in, we have
\begin{equation}
      \frac{2\widetilde{M}-A}{A^3} =\frac{\gamma^2\left(\beta\gamma - 1\right)}{\left(2-3\beta\gamma\right)^2}.
\end{equation}
We then indeed get the required
\begin{equation}
    k =  \kappa +\frac{\gamma^2\left(1-\beta\gamma \right)}{\left(2-3\beta\gamma\right)^2}.
\end{equation}
Plugging these all in to~\eqref{eq:Omega2BrA}, and considering $\widetilde{M}$ defined in~\eqref{eq:auxiliary}, we indeed find 
\begin{equation}\label{eq:CGS Final conf}
     \Omega^2 B(r) = 1-3 \beta \gamma-\frac{\beta(2-3 \beta \gamma)}{R}+\gamma R-\kappa R^2, 
\end{equation}
which is precisely $\widetilde{B}(R)$~\eqref{eq:CG Schwarzschild Lapse R}. We have thus shown that the CG Kerr metric~\eqref{eq:CG Kerr} is indeed conformally equivalent to CG Schwarzschild~\eqref{eq:CG Schwarzschild} in the zero spin limit.

When we apply the same conformal transformation to the full CG Kerr metric~\eqref{eq:CG Kerr} with a non-vanishing spin $a \neq 0$, we still have the coincidence of the four roots/horizons corresponding to the \textit{Empty} case. It may still be possible that another coordinate or conformal transformation may clarify this case, but we no longer pursue that here.

Furthermore, we note that we have an unphysical coordinate singularity when $\gamma = 0$ in the new coordinate system, which uses $R$ instead of $r$. The use of the coordinate $R$ instead of $r$ is not as well-adapted to the analysis of the geometry of the CG Kerr spacetimes as Boyer-Lindquist coordinates, as the $R$ coordinate of features now becomes directly dependent on $\gamma$ as from~\eqref{eq:R def} we have
\begin{equation}
    r = \frac{\left(2 - 3 \beta \gamma\right)R}{\gamma R + \left(2 - 3 \beta \gamma\right)}.
\end{equation}

A consequence of this is that features that may be radially ordered in $r$ may end up in unintuitive configurations. For example, when solving $\Delta^{\mathrm{H}} = 0$~\eqref{eq:rescaled} for $r$, given values of $\beta, \gamma, \kappa$ and $a$, we may have horizons ordered as $r^{\mathrm{H}}_1 < 0 < r^{\mathrm{H}}_2 < r^{\mathrm{H}}_3 < r^{\mathrm{H}}_4$. After we convert to the coordinate $R$, we find the horizons ordered as $R^{\mathrm{H}}_4 < R^{\mathrm{H}}_1 < 0 <  R^{\mathrm{H}}_2 < R^{\mathrm{H}}_3$. Some features may \textit{appear} on the opposite side of the singularity when we use $R$ instead of $r$. This issue is not encountered in the vanishing spin $(a = 0)$ limit, as the metric components cleanly reduce to the CG Schwarzschild form~\eqref{eq:CGS Final conf}.

\section{Derivation of extremal limit values}\label{section: Appendix quartics}

For the horizon extremal limits, we deal with the quartic $\Delta^\mathrm{H}$~\eqref{eq:rescaled}.
We consider the nature of roots of a general quartic equation of the form
\begin{equation}
c_4 r^4+c_3 r^3+c_2 r^2+c_1 r+c_0=0.
\end{equation}
From~\eqref{eq:rescaled}, we have $c_3 = 0$ and $c_2 = 1$. We thus have the discriminant

\begin{equation}
 D^\mathrm{H} \equiv  256c_4^3c_0^3 - 128c_4^2c_0^2 
+144c_4^2c_1^2c_0 - 27c_4^2c_1^4 + 16c_4c_0 - 4c_4c_1^2,
\end{equation}
giving~\eqref{eq:hordisc}, when we further consider that $c_4 = -k, c_1 = -2\widetilde{M},$ and $c_0 = a^2$.
We also have the associated quantities, where $k$ is as in~\eqref{eq:auxiliary}, given as
\begin{equation}
\begin{aligned}
&P\equiv 8c_4=-8k \\
& Z\equiv 64c_4^3c_0 - 16c_4^2=-16k^2\left(4k a^2+1\right) \\
& F \equiv 1 +12c_4c_0 = 1 - 12ka^2.
\end{aligned}
\end{equation} 
For four real roots, we require that $D^\mathrm{H} \geq 0$, $P < 0$, and $Z < 0$. 

When precisely $D^\mathrm{H} = 0$, two roots are repeated~\cite{quartic}, which is what we seek for an extremal limit. This condition is why we require~\eqref{eq:hordisc} to vanish.

For $\kappa \geq 0$, we automatically have $k \geq 0$ when considering $\gamma < 1$. Therefore, $P \leq 0, Z \leq 0$. In this case, we only have cosmological horizons when $D^{\mathrm{H}} < 0$.

Things are more complicated when $\kappa < 0$. When $D^{\mathrm{H}} < 0$, we either only have the two black hole horizons, or only a cosmological horizon. When $D^{\mathrm{H}} < 0$ and $P > 0$, which means $k < 0$, we only have the black hole horizons. When $D^{\mathrm{H}} < 0$ and $P < 0$, which means $k > 0$, we only have cosmological horizons.

Deriving the \textit{Empty} case for four repeated roots, we consider that all four roots of a general quartic are given by $-c_3/(4c_4)$ when $c_0/c_4 = 0$~\cite{quartic}. This yields
\begin{equation}
    \frac{(2-3\gamma)^2 a^2}{(2-3\gamma)^2\kappa+\gamma^2(1-\gamma)} = 0,
\end{equation}
which is satisfied when $\gamma = \frac{2}{3}$, giving the four repeated roots at $r = 0$ as we have found. We may be tempted to conclude that $a^2 = 0$ solves this. However, if $a^2 = 0$, $\Delta^\mathrm{H}$ reduces to $\Delta^\mathcal{E}$, which also has the \textit{Empty} case for $\gamma = \frac{2}{3}$ anyway.

Now, for the ergosurfaces, we factor out a power of $r$ and reduce $\Delta^\mathcal{E}$ to a cubic of the general form 
\begin{equation}\label{eq: ergo cubic}
 K_3 r^3 + K_2 r^2 + K_1 r + K_0 = -k r^3 + r  - (2-3  \gamma) = 0.
\end{equation}
In our case, $K_2 = 0$. The discriminant $ D^\mathcal{E}$ then simplifies to 
\begin{equation}
    D^\mathcal{E} \equiv-4K_3K_1^3 -27K_3^2K_0^2 = 4k -27k^2(2-3\gamma)^2 .
\end{equation}

A repeated root will be found when $D^\mathcal{E} = 0$~\cite{cubic}. Factoring out multiples of $ \left(2-3\gamma\right)$ and setting it to vanish gives us an equivalent to~\eqref{eq: ergo extremal gamma} of
\begin{equation}
 3 \gamma -27\kappa + 1  = 0.
\end{equation}

\clearpage

\section*{References}
\bibliography{Bibliography}

\end{document}